\documentclass[twocolumn, a4paper,accepted=2021-01-20]{quantumarticle}

\pdfoutput=1
\usepackage{subfiles}
\usepackage[latin9]{inputenc}
\usepackage[english]{babel}
\usepackage{amstext}
\usepackage{amssymb}
\usepackage{graphicx}
\usepackage{tabularx}
\usepackage{dsfont}
\usepackage{bbm}
\usepackage{caption}
\usepackage{subcaption}
\usepackage{amsmath}

\usepackage[numbers,sort&compress]{natbib}

\usepackage[usenames,dvipsnames,svgnames]{xcolor}
\definecolor{myurlcolor}{rgb}{0.2,0,0.8}
\definecolor{myrefcolor}{rgb}{0.2,0,0.8}
\usepackage[unicode=true, pdfusetitle,
            bookmarks=true, bookmarksnumbered=false,
            bookmarksopen=false, breaklinks=false,
            pdfborder={0 0 0}, backref=false,
            colorlinks=true, linkcolor=myrefcolor,
            citecolor=myurlcolor, urlcolor=myurlcolor]{hyperref}

\makeatletter

 \newcommand{\ket}[1]{| {#1} \rangle}
 \newcommand{\bra}[1]{\langle {#1}|}
 \newcommand{\ketbra}[2]{\ket{#1}\!\bra{#2}}
 \newcommand{\braket}[2]{\langle #1|#2\rangle}

\newcommand{\eins}{\mathbbm{1}}

\newcommand{\e}{{\mathrm{exp}}}

\newcommand{\E}{^{(\mathrm{e})}}
\newcommand{\B}{^{(\mathrm{b})}}

\newcommand{\K}[1]{\hat{\mathcal{K}}_{#1}^{\nu}}
\renewcommand{\L}[1]{\hat{\mathcal{L}}_{#1}^{\nu}}

\newcommand{\EPO}{$\langle\Box\rangle$} 

\renewcommand{\vec}[1]{\boldsymbol{#1}}

\renewcommand{\t}[1]{\textrm{#1}}

\newcommand{\eqnref}[1]{Eq.~(\ref{#1})}
\newcommand{\eqnsref}[2]{Eqs.~\eqref{#1} and \eqref{#2}}
\newcommand{\figref}[1]{Fig.~\ref{#1}}

\newcommand{\tabref}[1]{Table~\ref{#1}}
\newcommand{\secref}[1]{Sec.~\ref{#1}}
\newcommand{\appref}[1]{App. \ref{#1}}



\begin{document}

\title{A resource efficient approach for quantum and classical simulations of gauge theories in particle physics}

\author{Jan F. Haase}
\affiliation{Department of Physics \& Astronomy, University of Waterloo, Waterloo, ON, Canada, N2L 3G1}
\affiliation{Institute for Quantum Computing, University of Waterloo, Waterloo, ON, Canada, N2L 3G1}
\email{jan.frhaase@gmail.com}
\thanks{contributed equally}
\orcid{0000-0003-1126-8216}
\author{Luca Dellantonio}
\affiliation{Department of Physics \& Astronomy, University of Waterloo, Waterloo, ON, Canada, N2L 3G1}
\affiliation{Institute for Quantum Computing, University of Waterloo, Waterloo, ON, Canada, N2L 3G1}
\email{luca.dellantonio@uwaterloo.ca}
\thanks{contributed equally}
\orcid{0000-0003-1914-6539}
\author{Alessio Celi}
\affiliation{Departament de F\'isica, Universitat Aut\`onoma de Barcelona, E-08193 Bellaterra, Spain}
\affiliation{Center for Quantum Physics, Faculty of Mathematics, Computer Science and Physics, University of Innsbruck, Innsbruck A-6020, Austria}
\email{alessio.celi@uab.cat}
\thanks{contributed equally}
\orcid{0000-0003-4939-084X}
\author{Danny Paulson}
\affiliation{Department of Physics \& Astronomy, University of Waterloo, Waterloo, ON, Canada, N2L 3G1}
\affiliation{Institute for Quantum Computing, University of Waterloo, Waterloo, ON, Canada, N2L 3G1}
\orcid{0000-0003-2522-4046}
\author{Angus Kan}
\affiliation{Department of Physics \& Astronomy, University of Waterloo, Waterloo, ON, Canada, N2L 3G1}
\affiliation{Institute for Quantum Computing, University of Waterloo, Waterloo, ON, Canada, N2L 3G1}
\author{Karl Jansen}
\affiliation{NIC, DESY, Platanenallee 6, D-15738 Zeuthen, Germany}
\author{Christine A. Muschik}
\affiliation{Department of Physics \& Astronomy, University of Waterloo, Waterloo, ON, Canada, N2L 3G1}
\affiliation{Institute for Quantum Computing, University of Waterloo, Waterloo, ON, Canada, N2L 3G1}
\affiliation{Perimeter Institute for Theoretical Physics, Waterloo, ON, Canada, N2L 2Y5}
\orcid{0000-0002-4599-5107}

\begin{abstract}
	
Gauge theories establish the standard model of particle physics, and lattice gauge theory (LGT) calculations employing Markov Chain Monte Carlo (MCMC) methods have been pivotal in our understanding of fundamental interactions. 
The present limitations of MCMC techniques may be overcome by Hamiltonian-based simulations on classical or quantum devices, which further provide the potential to address questions that lay beyond the capabilities of the current approaches.
However, for continuous gauge groups, Hamiltonian-based formulations involve infinite-dimensional gauge degrees of freedom that can solely be handled by truncation. 
Current truncation schemes require dramatically increasing computational resources at small values of the bare couplings, where magnetic field effects become important. 
Such limitation precludes one from `taking the continuous limit' while working with finite resources.  
To overcome this limitation, we provide a resource-efficient protocol to simulate LGTs with continuous gauge groups in the Hamiltonian formulation. 
Our new method allows for calculations at arbitrary values of the bare coupling and lattice spacing. 
The approach consists of the combination of a Hilbert space truncation with a regularization of the gauge group, which permits an efficient description of the magnetically-dominated regime. 
We focus here on Abelian gauge theories and use $2+1$ dimensional quantum electrodynamics as a benchmark example to demonstrate this efficient framework to achieve the continuum limit in LGTs. 
This possibility is a key requirement to make quantitative predictions at the field theory level and offers the long-term perspective to utilise quantum simulations to compute physically meaningful quantities in regimes that are precluded to quantum Monte Carlo.

\end{abstract}

\tableofcontents{}

%
%

\section{Introduction}

Gauge theories are the basis of high energy physics and the foundation of the standard model (SM). 
They describe the elementary interactions between particles, which are mediated by the electroweak and strong forces \cite{Cottingham2007An-Introduction, Altarelli2017Collider, Peskin1995ev}, making the SM one of the most successful theories with tremendous predictive power \cite{Altarelli2014The-Higgs}. 
Still, there are numerous phenomena which cannot be explained by the SM. 
Examples include the nature of dark matter, the hierarchy of forces and quark masses, the matter antimatter asymmetry and the amount of CP violation \cite{Veltman:2018nzg}. 
Answering these questions and accessing physics beyond the SM, though, often requires the study of non-perturbative effects.

A very successful approach to address non-pertubative phenomena is lattice gauge theory (LGT) \cite{DeGrand:2006zz,Rothe1992Lattice,Gattringer:2010zz}, as proposed by Kenneth Wilson in 1974 \cite{Wilson1974Confinement}. 
In LGT, Feynman's path integral formulation of quantum field theories (QFTs) is employed on an Euclidean space-time grid. 
Such a discretised form of the path integral allows for numerical simulations utilizing Markov Chain Monte Carlo (MCMC) methods. 
The prime target of LGT is quantum chromodynamics (QCD), i.e. the theory of strong interactions between quarks and gluons. 
In this field, LGT has been extremely successful, allowing for example the computation of the the low-lying baryon spectrum \cite{Durr:2008zz}, the structure of hadrons, fundamental parameters of the theory and many more \cite{Constantinou:2015agp,Cichy:2018mum,MEYER201946,Juettner:2016atf}.

However, many of the aforementioned open questions in modern physics cannot be addressed within the standard approach, due to the sign-problem \cite{Troyer:2004ge,Banuls:2019rao,Banuls:2016gid} that renders MCMC methods ineffective. 
A possible solution is to employ a Hamiltonian formulation of the underlying model. 
Classical Hamiltonian-based simulations using tensor network states (TNS),
including fermionic projected entangled-pair states, have been successful \cite{Sugihara2005Matrix,Tagliacozzo2014Tensor,Rico2014,Zohar2015Fermionic, Haegeman2015Gauging, Pichler2016,Zohar2016Building, Silvi2017,Banuls2018Tensor, Magnifico2019,Chanda2020Confinement}, but are so far restricted to mostly one spatial dimension (for link model 2D calculations with DMRG and tree tensor network see e.g \cite{Tschirsich19,Felser2020}). 
Consequently, there is a necessity for new approaches to both access higher dimensions and address problems where standard MCMC methods fail.
It is presently not known whether efficient classical methods can be developed to overcome this problem.

\begin{table*}[t]
\caption{\textbf{Computational cost for different approaches.} We estimate the number of states required to reach a $1\%$ accuracy in the expectation value of the two-dimensional plaquette in QED (see Sec. 4.3) when compared to the value we obtain with our method considering a maximum of $9261$ basis elements. The three columns refer, from left to right, to the standard approach described in \secref{sec:QEDin2D}, our approach (see \secref{sec:transformation}) using a fixed group $\mathbb{Z}_N$, and finally our optimised strategy, in which the order of the group $N$ is scaled with the bare coupling $g$ (see \secref{sec:performance}). The shown savings in computational resources bring quantum simulations with current technology within reach. Note that $125$ states correspond to seven qubits. We present a robust implementation strategy for ion-based quantum computers in \cite{Papero2}.}

\label{tab:num_states}
\centering
\includegraphics[scale=0.25]{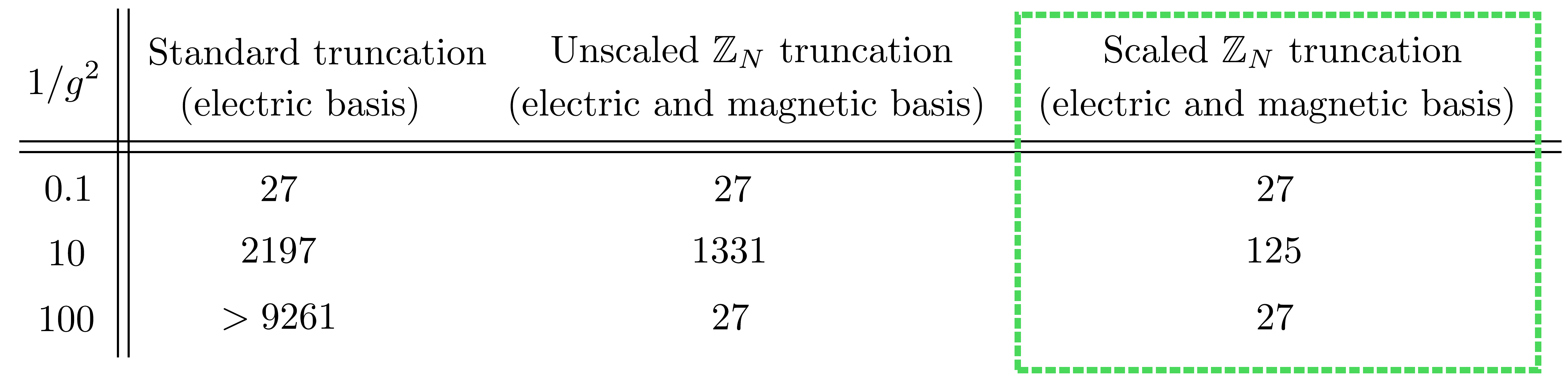}
\end{table*}

Hamiltonian-based simulations on quantum hardware provide an alternative route, since there is no such fundamental obstacle to simulating QFTs in higher dimensions \cite{Byrnes2006Simulating, Preskill2018Simulating, Banuls2019Simulating, Banuls2019Review}.
Therefore, this approach holds the potential to address questions that cannot be answered with current and even future classical computers. 
The rapidly evolving experimental capabilities of quantum technologies \cite{Schleich2016Quantum, Acin2018The-quantum} have led to proof-of-concept demonstrations of simulators tackling one-dimensional theories \cite{Martinez2016Real-time, Klco2018Quantum-classical, Kokail2019Self-verifying, Mil2020A-scalable, Yang2020Observation, Schweizer2019Floquet, Gorg2019Realization}. 
Extending these results to higher dimensions is a lively area of research \cite{Byrnes2006Simulating,Weimer2010rydberg, Zohar2012Simulating,Tagliacozzo2013Optical, Tagliacozzo2013Simulation, Glaetzle2014Quantum, Dutta2017Toolbox, Zohar2017Digital, Hackett2019Digitizing, Lamm2019General, celi2019emerging}, since it represents a crucial step for this field, and realisations on `Noisy Intermediate-Scale Quantum' devices \cite{Georgescu2014Quantum, Preskill2018Quantum}, i.e. current quantum hardware, require novel approaches to make this leap.

To meet this challenge, we provide a resource-efficient approach that facilitates the quantum simulation of higher dimensional LGTs that would otherwise be out of reach for current and near-term quantum hardware, which is exemplified \tabref{tab:num_states}. 
In addition, purely classical simulations based on the Hamiltonian formalism also benefit from our resource-optimised approach. 
Hence, we bring both quantum and classical calculations closer to developing computational strategies that do not rely on Monte Carlo methods, and thus circumvent their fundamental limitations.

Our new approach addresses the important problem of reaching the continuum limit (in which the lattice spacing approaches zero) with finite computational resources. 
Since QFTs are continuous in their time and space variables, the need to take a controlled continuum limit is inherent to any lattice approach and necessary to extract physically relevant results from a lattice simulation.

Taking QCD as a concrete example, we require an accurate description for particles interacting at both short and long distances. 
Lattice QCD and other LGTs offer the unique tool to investigate both regimes. 
At long distances, e.g. the bound state spectrum can be computed. 
At short distances, and after taking the continuum limit, it is possible to connect the perturbative results derived with QFTs with non-perturbative simulations, thus assessing the range in which perturbation theory is valid. 
However, taking the continuum limit is in general computationally expensive. 
MCMC methods, for instance, have the intrinsic problem of autocorrelations, that become more and more severe when decreasing the lattice spacing. 
This drawback in turn leads to a significant increase in the computational cost, and fixes the smallest value of the lattice spacing that can be reached. 
On one hand, Hamiltonian approaches circumvent this problem. 
On the other, however, Hamiltonian-based formulations face the challenge that for continuous (Abelian and non-Abelian) gauge groups, local gauge degrees of freedom are defined in an infinite dimensional Hilbert space. 
As a consequence, any simulation -- classical or quantum -- requires a truncation of the gauge fields, which is inherently at conflict with the required continuum limit.

In this work, we present a practical solution to overcome this crucial bottleneck and to allow for resource-efficient Hamiltonian simulations of LGTs. 
Although our approach is general and applicable to LGTs of any dimension, we consider two-dimensional quantum electrodynamics (QED) as a benchmark example.

Truncation of the gauge fields is typically performed in the `electric basis', i.e. the basis in which the electric Hamiltonian and Gauss' law are diagonal. 
As such, truncation preserves the gauge symmetry, and the resulting models are known as gauge magnets or link models \cite{Horn1981Finite,Orland1990Lattice,Chandrasekharan1997Quantum}, which are of direct relevance in condensed matter physics \cite{Rokhsar1988Superconductivity,Moessner2001Short-ranged,Fradkin2013Field,Sachdev2016Emergent}. 
As recently shown in Ref. \cite{celi2019emerging}, spin-$1/2$ truncations are within the reach of current quantum simulators. 
From the perspective of fundamental particle interactions, electric truncations can result in an accurate description of the system in the strong coupling regime.
 However, by decreasing the value of the coupling or equivalently the lattice spacing, the magnetic contributions to the energy become increasingly important and the number of states that have to be included in the electric basis grows dramatically (a similar increase can be realised by adding an auxiliary spatial dimension to the lattice \cite{PhysRevD.60.094502}). 
An alternative approach to describe the gauge degrees of freedom is to approximate continuous gauge groups with discrete ones, for instance, to approximate $U(1)$ with $\mathbb{Z}_{2L+1}$ ($L \in \mathbb{N}$) \cite{Zohar2013Quantum, Kuhn2014Quantum, Ercolessi2018Phase}. 
Such approaches also face similar limitations as the ones described above, as $L$ has to be progressively increased with decreasing coupling.

A natural solution to simulate the weak coupling regime consists of exploiting the self-duality \cite{Kogut1983The-lattice} of the electric and magnetic terms by Fourier transforming the Hamiltonian and working in the `magnetic basis', i.e. the basis in which magnetic interactions are diagonal, as suggested in \cite{stryker2018gauss}.
However, the fact that the magnetic degrees of freedom are continuous variables with a gapless spectrum, poses intricate challenges for a resource-efficient truncation scheme, that have yet (to the best of our knowledge) to be addressed. 
In this work, we provide a practical solution by combining state truncation with a gauge group discretisation that is dynamically adjusted to the value of the coupling. 
This approach allows for controlled simulations at all values of the bare coupling, smoothly connecting the weak, strong and intermediate coupling regimes. 
As a proof-of-principle of this new approach and its ability to faithfully simulate non-perturbative phenomena, we target the renormalised coupling in QED in $2 + 1$ dimensions.

To observe non-perturbative phenomena such as confinement, the simulated physical length scale needs to be larger than the scale at which confinement sets in. 
As a result, large lattice sizes are required and the number of lattice points grows rapidly when approaching the continuum limit of the theory. 
This results in computational requirements that cannot be satisfied using current classical and quantum computers. 
Still, as previously done in the pioneering work by Creutz \cite{Creutz:1980zw}, we can study the bare coupling dependence of the local plaquette operator. 
This quantity allows us to benchmark our formalism and to show that a smooth connection between the weak and the strong coupling regimes can be established. 
In addition, our method allows for estimating the precision with which a given truncation approximates the untruncated results.

The paper is organised as follows. 
In \secref{sec:2}, we review lattice QED in $2+1$ dimensions as an example of Abelian and non-Abelian gauge theories with continuous groups and magnetic interactions. 
We consider lattices with periodic boundary conditions and reformulate the lattice Hamiltonian in terms of gauge-invariant degrees of freedom. 
By eliminating redundant variables, we obtain an effective Hamiltonian description that allows for simulations at a low computational cost. 
In \secref{sec:transformation}, we introduce a new magnetic representation of lattice QED that is equipped with a regularisation in terms of a $\mathbb{Z}_{2L+1}$ group and an efficient truncation scheme. 
In \secref{sec:performance}, we study the performance of our method and benchmark its precision by calculating the expectation value of the plaquette operator on a periodic plaquette in the static charge limit. 
We show that both the truncation cut-off parameter, i.e. the maximum number of gauge basis states included in the simulation, and $L$, the dimension of the $\mathbb{Z}_{2L+1}$ group, can be used as adjustable variational parameters. Both are used to optimise the simulation and estimate its accuracy. 
In the following \secref{sec:Generalisations}, we present the generalisation to an arbitrary, two-dimensional periodic lattice with dynamical matter. Finally, we outline the prospects of this method for classical and quantum simulations is in \secref{sec:Conclusions}

\section{Minimal encoding of LGTs with continuous gauge groups}
\label{sec:2}
In this chapter, we provide a Hamiltonian formulation for LGTs with continuous gauge groups that allows for resource-efficient classical and quantum simulations. 
First, we review the standard Kogut-Susskind Hamiltonian subject to Gauss' law (the local constraints ensuring gauge invariance) in \secref{sec:QEDin2D}, considering QED on a square lattice as a paradigmatic example. 
In \secref{sec:single_plaq_ham}, we proceed to provide a minimal formulation of the QED lattice Hamiltonian, in which redundant degrees of freedom have been removed.

\subsection{QED in two dimensions}
\label{sec:QEDin2D}

We review here the bottom-up construction of the lattice Hamiltonian as originally presented in \cite{Kogut1975Hamiltonian}. 
For the sake of simplicity, we consider QED in $2+1$ dimensions which displays key features of phenomenologically relevant theories like QCD, including chiral symmetry breaking and a renormalisation of the coupling constant \cite{raviv2014nonperturbative}, features that 
are absent in one spatial dimension. 

The Hamiltonian of Abelian and non-Abelian gauge theories in two (or more) dimensions is constructed in terms of electric and magnetic fields, and their coupling to charges. 
In continuous Abelian $U(1)$ gauge theories like QED (and similarly for non-Abelian gauge theories like QCD), electric and magnetic fields are defined through the vector potential $A_\mu$, with $E_\mu=\partial_t A_\mu$ and $B=\partial_x A_y-\partial_y A_x$ (in the unitary gauge $A_0=0$). 
Here $t,x,y$ are the time and space coordinate in two dimensions, and $\mu=x,y$.

Gauge invariance, i.e. invariance of the Hamiltonian under local phase (symmetry) transformations of the charges, follows directly from the invariance of $E_\mu$ and $B$ under $A_\mu\to A_\mu + \partial_\mu \theta(x,y)$, where $\theta(x,y)$ is an arbitrary scalar function.
Due to the unitary gauge, only spatial, time-independent transformations are considered.
The electric field is sourced by the charges through Gauss' law, $\sum_\mu\partial_\mu E_\mu =4 \pi \rho$, where $\rho$ is the charge density. 

\begin{figure*}[ht!]
\includegraphics[width=2\columnwidth]{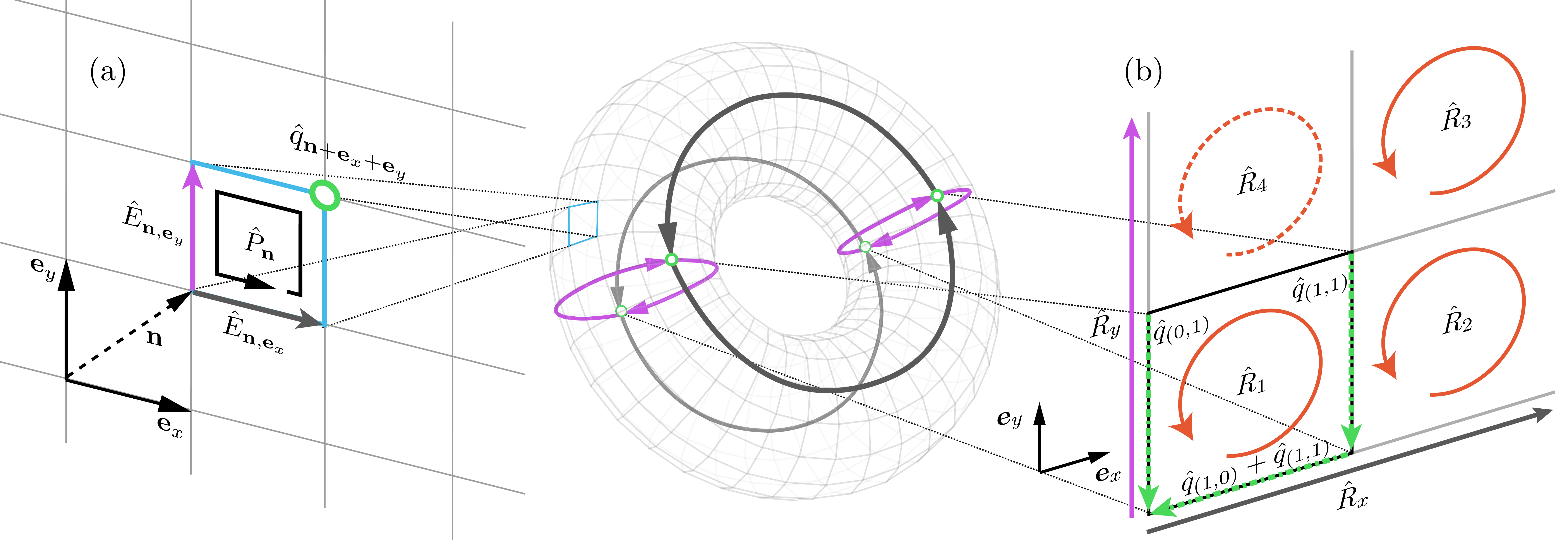}
\caption{\textbf{Two-dimensional lattice gauge theory with periodic boundary conditions.} A single cell of the periodic 2D lattice in (a) is made of four links, oriented towards the positive $x$ and $y$ directions. Each lattice site is indicated by a unique vector $\vec{n}$, which marks the lower left corner of each single plaquette. The associated operator $\hat{P}_{\vec{n}}$ accounts for the electric field quanta circulating along the sketched path. The periodic lattice spans the surface of a torus, shown in the middle, whose minimal instance is assembled by four sites and the corresponding electric fields [thick lines, same color coding as in (a)]. Unwrapping this minimal torus yields the geometry shown in (b). We identify the strings $\hat{R}_{x}$ and $\hat{R}_{y}$ and the four rotators $\hat{R}_j$, $j=1,2,3,4$. The eigenstates of the strings and three of the rotators (we arbitrarily remove $\hat{R}_4$, dashed loop) form a basis for the physical states of the pure gauge theory. To describe the physical states for a generic charge configuration we add three charge strings (dotted green arrows) that correspond to a conventional physical state for the given charge configuration.}
\label{fig:theory}
\end{figure*}

In LGTs \cite{Wilson1974Confinement}, the charges occupy the sites $\vec{n} =(n_x,n_y)$ of the lattice while the electromagnetic fields are defined on the links. 
The links are denoted by their starting site $\vec{n}$ and their direction $\vec{e}_{\mu}$ ($\mu=x,y$), as shown in \figref{fig:theory}. 
The electric interactions are defined in terms of the electric field operator $\hat{E}_{\vec{n},\vec{e}_{\mu}}$, which is Hermitian, possesses a discrete spectrum and acts on the link connecting the sites with coordinates $\vec{n}$ and $\vec{n}+\vec{e}_\mu$. 
For each link, one further defines a Wilson operator  $\hat U_{\vec{n},\vec{e}_{\mu}}$, as the lowering operator for the electric field, $[\hat{E}_{\vec{n},\vec{e}_\mu},\hat{U}_{\vec{n}^\prime,\vec{e}_\nu}] = - \delta_{\vec{n},\vec{n}^\prime} \delta_{\mu,\nu}\hat{U}_{\vec{n},\vec{e}_\mu}$.
The Wilson operator measures the phase proportional to the bare coupling $g$ acquired by a unit charge moved along the link $(\vec{n},\vec{e}_\mu)$ of length $a$, i.e. $\hat U_{\vec{n},\vec{e}_\mu} \sim \exp\lbrace i a g\hat{A}_\mu(\vec{n})\rbrace$.
The magnetic interactions are given by (oriented) products of Wilson operators
on the links around the plaquettes of the lattice.
These operators are used to construct the Kogut-Susskind Hamiltonian as $\hat H = \hat H_\text{gauge} + \hat H_\text{matter}$. 
Let us discuss first the pure gauge part that describes the limit of static charges 
\begin{eqnarray}\label{eq:HKS}
\hat{H}_\text{gauge} &=& \hat{H}_{E} + \hat{H}_{B},\nonumber\\	
\hat{H}_E &=& \frac{g^{2}}{2} \sum_{\vec{n}} \left(\hat{E}^{2}_{\vec{n},\vec{e}_x} + \hat{E}^{2}_{\vec{n},\vec{e}_y}\right),\nonumber\\ 
\hat{H}_B &=& -\frac{1}{2g^{2}a^2} \sum_{\vec{n}} \left(\hat{P}_{\vec{n}} + \hat{P}_{\vec{n}}^{\dag}\right). 
\end{eqnarray}
Here,  the sums run over both components of the sites $\vec{n} = (n_x,n_y)$ and
\begin{eqnarray}\label{eq:PlaquetteOpDef}
\hat{P}_{\vec{n}} =  \hat{U}_{\vec{n},\vec{e}_x}\hat{U}_{\vec{n}+\vec{e}_x,\vec{e}_y}\hat{U}^{\dag}_{\vec{n}+\vec{e}_y,\vec{e}_x}\hat{U}^{\dag}_{\vec{n},\vec{e}_y}
\end{eqnarray}
is the plaquette operator.
It is easy to check that \eqnref{eq:HKS} reduces to the pure gauge $U(1)$ Hamiltonian in the continuum, $\hat{H}\propto\int \mathrm{d}\vec{x} E(\vec{x})^2 + B(\vec{x})^2$, when the lattice spacing $a$ is sent to zero (see \appref{app:dimensions}). 
The Hamiltonian in \eqnref{eq:HKS} is gauge-invariant as it commutes with the lattice version of Gauss' law
\begin{eqnarray}\label{eq:GaussKS}
 \Bigg[\sum_{\mu=x,y}&&
\left(\hat{E}_{\vec{n},\vec{e}_\mu} -\hat{E}_{\vec{n}-\vec{e}_\mu,\vec{e}_\mu} \right) - \hat{q}_{\vec{n}} - \hat{Q}_{\vec{n}}\Bigg] \ket{\Phi} = 0  \; \forall \vec{n} \nonumber\\
&& \iff \ket{\Phi} \in \lbrace\text{physical states}\rbrace,
\end{eqnarray}
that determines what states are physical for a given distribution of charges. Here, $\hat{q}_{\vec{n}}$ is the operator measuring the charge on the site $\vec{n}$ and 
$\ket{\Phi}$ represents the state of the whole lattice, including both links and sites. 
Furthermore, the operators $\hat{Q}_{\vec{n}}$ denote possible static charges which we set to zero in the following.
The eigenstates of the electric field operators 
\begin{eqnarray}
\hat{E}_{\vec{n},\vec{e}_\mu} \ket{E_{\vec{n},\vec{e}_\mu}} = E_{\vec{n},\vec{e}_\mu} \ket{E_{\vec{n},\vec{e}_\mu}}, \quad E_{\vec{n},\vec{e}_\mu} \in \mathbb{Z}
\end{eqnarray}
form a basis for the link degrees of freedom.
In particular, the physical states can be easily identified in this basis via \eqnref{eq:GaussKS}.

Let us now consider moving charges. 
To ensure gauge invariance, their motion is required to respect Gauss' law, i.e. 
a charge $q$ moving between two sites has to change the electric field along the path by $-q$. 
In other words, the lowering operator $\hat{U}$ has to be applied $q$ times to the links on the path to preserve gauge-invariance.
Since $\hat{U}^q = \exp \lbrace i q a g \hat{A} \rbrace$, the so-called minimal coupling condition \cite{Rothe1992Lattice} is recovered in the continuum limit $a \rightarrow 0$, which is equivalent to replacing derivatives of matter fields by the covariant derivatives, i.e. shifting the particles' momentum by a gauge field contribution $\hat{p}_\mu \mapsto \hat{p}_\mu - q g \hat{A}_\mu$.

In QED, charges are represented by Dirac fermions.
In the staggered representation \cite{Kogut1975Hamiltonian}, 
they are represented on a square lattice as ordinary fermions at half filling, with staggered chemical potential that plays the role of the mass term.
Their Hamiltonian is $\hat{H}_\t{matter} = \hat{H}_{M} + \hat{H}_{K}$, where $\hat{H}_{M}$ and $\hat{H}_{K}$ are the mass and kinetic contributions, respectively
\begin{eqnarray}
\hat{H}_{M} &=& m\sum_{\vec{n}} (-1)^{n_x+n_y} \hat{\Psi}^\dag_{\vec{n}} \hat{\Psi}_{\vec{n}}, \label{eq:MassContrGen} \\
\hat{H}_{K} &=& \kappa \sum_{\vec{n}} \sum_{\mu = x,y} \left[\hat{\Psi}_{\vec{n}}^{\dag} \left(\hat{U}_{\vec{n},\vec{e}_\mu}^{\dagger}\right)^q \hat{\Psi}_{\vec{n}+\vec{e}_\mu} + \t{H.c.}\right]. \label{eq:KineticContrGen}
\end{eqnarray}
Here, $m$ and $q$ are the particles' effective mass and (integer) charge, $\kappa$ the kinetic strength and $\hat{\Psi}_{\vec{n}}^{(\dag)}$ the fermionic lowering (raising) operator for site $\vec{n}$. 
Since $\hat H_M$ identifies the Dirac vacuum with the state with all odd sites occupied, creating (destroying) a particle at even (odd) site is equivalent to creating a $(-)q$-charged ``fermion'' (``antifermion'') in the Dirac vacuum. 
Thus the tunneling processes in the kinetic term correspond to the creation or annihilation of particle-antiparticle pairs and the corresponding change in the electric field string connecting them. 
The charge operator $\hat{q}_{\vec{n}}$ is given by
\begin{eqnarray}\label{eq:charge_op}
\hat{q}_{\bf{n}} = q \left( \hat{\Psi}_{\bf{n}}^{\dag}\hat{\Psi}_{\bf{n}} - \frac{\mathbbm{1}}{2}[1-(-1)^{n_x+n_y}] \right),
\end{eqnarray}
where $q$  is an integer number which we set to one in the following.
Note that we rescaled the fermion field by a factor $\sqrt{\alpha}$ as discussed in \appref{app:dimensions}, which establishes the relations 
\begin{eqnarray}
m = \frac{M}{\alpha}\;\; \t{and} \;\; \kappa = \frac{1}{2a\alpha},
\end{eqnarray}
with $M$ being the bare mass of the particles.

We conclude this section with a few comments on the structure of the pure gauge part of the Kogut-Susskind Hamiltonian in \eqnref{eq:HKS}. 
There, the electric and magnetic terms show an apparent asymmetry that obscures the electromagnetic duality in QED in the continuum and in the Wilson lattice formulation \cite{Wilson1974Confinement}. 

The symmetry between electric and magnetic fields in QED and in Wilson's action-formulation is due to the fact that time and space are treated on the same footing. 
Wilson's lattice action theory is formulated on a space-time grid with lattice spacing $a_\mu$, $\mu = t,x,y,z$. 
In this case, an isotropic continuum limit is taken in which the lattice spacings in both the temporal and the spatial directions approach zero. 
In the Hamiltonian formulation, time is a continuous parameter. 
Accordingly, the above procedure is broken into two steps. 
Firstly, the continuum limit with respect to time $a_t \rightarrow 0$ is taken, which results in the Hamiltonian lattice formulation used here. 
In a second step, the continuum limit has to be taken with respect to space $a_{x,y,z} \rightarrow 0$ to obtain physical results.

In the Hamiltonian formulation, the electric field operators $\hat{E}$ form an algebra and are non-compact, as their integer spectrum takes values from minus infinity to infinity. 
By contrast, the Wilson operators $\hat{U}$ and hence the plaquette operators $\hat{P}$, form a group. 
The moduli of their expectation value is one, as is the case for the Wilson action. 
More specifically, in Wilson's action formulation, it follows from operators $\hat{U}=\exp\lbrace i g a_\mu \hat{A}_\mu\rbrace$ that  $\hat{A}_\mu$ is \emph{compact} as it is defined from $-\pi/(g a_\mu)$ to $\pi/(g a_\mu)$, where $a_\mu$ is the lattice spacing in the $\mu$-direction, which includes both space and time as $\mu = t,x,y,z$. 
The Kogut-Susskind Hamiltonian can be obtained from the Wilson action by taking the continuum limit in the time direction $a_t \rightarrow 0$ \cite{Creutz:1980zw}. 
Thus, the asymmetry between the electric and magnetic terms in the Hamiltonian formulation disappears when the continuum limit is taken in the spatial direction.

While a fully non-compact formulation of Hamiltonian LGT is possible \cite{Drell1979Quantum} (for the different outcomes of the two approaches see e.g. \cite{Fiebig1990Monopoles, Herbut2003Permnent}), we do not discuss this approach here as it is not advantageous for quantum simulations. 
As we show in \secref{sec:transformation}, it is instead convenient to write the electric term in a compact form.

%
%

\subsection{QED Hamiltonian for physical states}
\label{sec:single_plaq_ham}

As outlined in the previous section, gauge-invariance constrains the dynamics to the physical states only, i.e. those satisfying Gauss' law in \eqnref{eq:GaussKS}. 
Practically, unphysical states have to be suppressed, e.g. via energy penalties \cite{Halimeh2020Gauge-Symmetry}.
In any case, quantum states that are not physical represent an exponential overhead for classical and quantum computation (also after a proper truncation, see \secref{sec:performance}). 
Furthermore, in noisy near term quantum devices or simulation protocols where the Hamiltonian has to be split up, e.g. to simulate time-evolutions employing a product formulas such as the Trotter expansion \cite{Raedt1987Product}, implementing or imposing Gauss' law during the simulation may be complicated, or even impossible. 

It is thus convenient to eliminate the redundant degrees of freedom by solving the constraint at each lattice site. 
In one dimension, such a procedure allows one to completely eliminate the gauge field, yielding an effective Hamiltonian containing only matter terms (but long-range interactions) \cite{Hamer1997, Muschik2017U1-Wilson}.  
A similar approach is applicable in higher dimensions, with the difference that the gauge field has also physical degrees of freedom. 
Here, we show how to formulate an effective Hamiltonian that directly incorporates the constraints of \eqnref{eq:GaussKS} 
by employing a convenient parametrization of the physical states that yields an intuitive description of the system.

For the sake of clarity, we consider the minimal instance of a periodic two-dimensional gauge theory: a square lattice formed by four lattice points.
The generalisation to an arbitrary lattice on a torus is derived in \secref{sec:Generalisations}.
Due to the periodic boundary conditions, this minimal system can equivalently be represented as a torus with four faces, or as four distinct plaquettes consisting of eight links [see \figref{fig:theory}(b)]. 
Due to charge conservation $\sum_{\vec{n}}\hat{q}_{\vec{n}}=0$, only three out of the four constraints given by Gauss' law [\eqnref{eq:GaussKS}] are independent. 
Consequently, three of the eight links in the lattice are redundant, and the electric Hamiltonian in \eqnref{eq:HKS} can be solely expressed in terms of the remaining five (see \appref{app:Reparametrisation} for details).

Describing the system in terms of these five links, however, entails serious drawbacks. 
The effective Hamiltonian contains many body interactions which are challenging or even impossible to be implemented using available quantum hardware (see \appref{app:Reparametrisation}). 
To circumvent this problem, we consider a natural basis for the physical states in terms of small loops around each plaquette, and large electric loops around the whole lattice.  
In such a basis, the electric and magnetic interactions take a simple form. 
To conveniently describe these interactions, we introduce a set of operators, \textit{rotators} and \textit{strings} (see Fig. \ref{fig:theory}), that are diagonal and label the loop basis. 
As we show in \cite{Papero2}, the Hamiltonian formulated in terms of these operators can be simulated with current quantum hardware.

With the notation and conventions presented in \figref{fig:theory}, rotators and strings are given by the relations
\begin{eqnarray}\label{eq:gaugefixing}
\hat{E}_{(0,0),\vec{e}_{x}} &=& \hat{R}_1 + \hat{R}_x -(\hat{q}_{(1,0)}+\hat{q}_{(1,1)}), \nonumber \\
\hat{E}_{(1,0),\vec{e}_{x}} &=& \hat{R}_{2} - \hat{R}_{3} + \hat{R}_x ,\nonumber \\
\hat{E}_{(1,0),\vec{e}_{y}} &=& \hat{R}_1 - \hat{R}_{2} -\hat{q}_{(1,1)} ,\nonumber \\
\hat{E}_{(1,1),\vec{e}_{y}} &=& -\hat{R}_{3},\nonumber \\
\hat{E}_{(0,1),\vec{e}_{x}} &=& -\hat{R}_1,\nonumber \\
\hat{E}_{(1,1),\vec{e}_{x}} &=& \hat{R}_{3} - \hat{R}_{2},\nonumber \\
\hat{E}_{(0,0),\vec{e}_{y}} &=& \hat{R}_{2} - \hat{R}_1 + \hat{R}_y - \hat{q}_{(0,1)},\nonumber \\
\hat{E}_{(0,1),\vec{e}_{y}} &=& \hat{R}_{3} + \hat{R}_y,
\end{eqnarray}
where the charges $\hat{q}_{\vec{n}}$ are required by Gauss' law. 
An intuitive way to understand the effect of the charge terms in \eqnref{eq:gaugefixing} is to consider them as sources of additional electric strings (whose concrete choice is just a matter of convention but consistent with Gauss' law), as displayed by the green lines in \figref{fig:theory}(b). We remark that this becomes evident from the link formulation in \appref{app:Reparametrisation}. In \appref{app:intuitive} we give an alternative explanation of the form of \eqnref{eq:gaugefixing}. 

As mentioned above, rotators and strings automatically preserve Gauss' law, which can be readily verified by observing that at any site, the incoming fields are always balanced by the outgoing ones. 
Moreover, by recalling the plaquette operator $\hat{P}_{\vec{n}}$ in \eqnref{eq:PlaquetteOpDef}, it becomes clear why $\hat{R}_{i}$ and $\hat{R}_{\mu}$ are a convenient choice to represent the electric gauge field components. 
The operator $\hat{P}_{\vec{n}}$ increases the anticlockwise quanta of the electric field circulating in the $\vec{n}$-th plaquette. 
Consequently, it does not act on strings and takes the form of the lowering operator of the associated rotator. 
This fact can be formally proven by examining the raising and lowering operators of rotators and strings. 
From the commutation relations of the links and the relations shown in \eqnref{eq:gaugefixing}, it follows that
\begin{align}\label{eq:rotatorstringcommrel}
\big[ \hat{R}_{i},\hat{P}_{j} \big] =&  \delta_{i,j} \hat{P}_{j},\nonumber \\
\big[ \hat{R}_{x} , \hat{U}_{(0,0),\vec{e}_{x}}\hat{U}_{(1,0),\vec{e}_{x}} \big] =& \hat{U}_{(0,0),\vec{e}_{x}}\hat{U}_{(1,0),\vec{e}_{x}} \equiv \hat{P}_{x}, \nonumber \\
\big[ \hat{R}_{y} , \hat{U}_{(0,0),\vec{e}_{y}}\hat{U}_{(0,1),\vec{e}_{y}} \big] =& \hat{U}_{(0,0),\vec{e}_{y}}\hat{U}_{(0,1),\vec{e}_{y}} \equiv \hat{P}_{y},
\end{align}
where $\hat{P}_{j}$, $j=1,2,3$ is the plaquette operator of plaquette $j$ as denoted in \figref{fig:theory}. 
Moreover, we defined the string lowering operators $\hat{P}_{x} \equiv \hat{U}_{(0,0),\vec{e}_{x}}\hat{U}_{(1,0),\vec{e}_{x}}$ and $\hat{P}_{y} \equiv \hat{U}_{(0,0),\vec{e}_{y}}\hat{U}_{(0,1),\vec{e}_{y}}$. 

The magnetic Hamiltonian for the periodic plaquette in \figref{fig:theory}(b),
\begin{equation}
\hat{H}_{B} = -\frac{1}{2 g^2 a^2} \left(\hat{P}_{1} + \hat{P}_{2} + \hat{P}_{3} + \hat{P}_{4} + H.c.\right),
\end{equation}
is proportional to the sum of four plaquette operators, while there are only three independent rotators. 
The fourth rotator can be written as a combination of the others, since the effect of lowering (raising) all other rotators, i.e. $\hat{R}_{1}$, $\hat{R}_{2}$ and $\hat{R}_{3}$, amounts to raising (lowering) $\hat{R}_{4}$. 
This relation can be understood by examining \eqnref{eq:gaugefixing}: By lowering all of the three rotators $\hat{R}_{1}$, $\hat{R}_{2}$ and $\hat{R}_{3}$, we manipulate the electric fields on the links constituting $\hat{R}_{4}$ in exactly the same way as an increment of the latter would do. 
As such, the magnetic Hamiltonian becomes
\begin{equation}\label{eq:MagneticContrSingle}
\hat{H}_{B} = -\frac{1}{2 g^2 a^2} \left( \hat{P}_{1} + \hat{P}_{2} + \hat{P}_{3} + \hat{P}_{3}^{\dag}\hat{P}_{2}^{\dag}\hat{P}_{1}^{\dag} + \t{H.c.} \right),
\end{equation}
while, by inserting \eqnref{eq:gaugefixing} into \eqnref{eq:HKS}, the electric term takes the form:
\begin{align}\label{eq:ElectricContrSingle}
\hat{H}_E =&\, g^2\Bigg \lbrace2\left[\hat{R}_1^2  + \hat{R}_2^2 + \hat{R}_{3}^2 - \hat{R}_{2}(\hat{R}_{1} + \hat{R}_{3})\right] \nonumber\\
&+  \hat{R}_x^2 + \hat{R}_y^2 + (\hat{R}_1 + \hat{R}_2 - \hat{R}_{3}) \hat{R}_x   \nonumber\\
&-  (\hat{R}_1 - \hat{R}_2 - \hat{R}_{3}) \hat{R}_y  \nonumber \\
&-  \left[  \hat{q}_{(1,0)} (\hat{R}_1 + \hat{R}_x) \right. \nonumber \\
&+  \left.  \hat{q}_{(0,1)} (\hat{R}_{2} - \hat{R}_1  +\hat{R}_y)\right.  \nonumber\\
&+  \left. \hat{q}_{(1,1)} ( 2 \hat{R}_1 - \hat{R}_{2} + \hat{R}_x)  \right] \nonumber \\
&+ . \frac{\hat{q}_{(1,0)}^2 +  \hat{q}_{(0,1)}^2 + 2\hat{q}_{(1,1)}(\hat{q}_{(1,0)} +\hat{q}_{(1,1)})}2
\Bigg \rbrace.
\end{align}

Once the effective gauge Hamiltonian $\hat{H}_{\rm gauge} = \hat{H}_{E}+\hat{H}_{B}$ has been derived in terms of rotator and string operators, we must further modify the matter Hamiltonian $\hat{H}_{\rm matter} = \hat{H}_{M}+\hat{H}_{K}$ [for the description in terms of field operators, see \appref{app:Reparametrisation}]. 
While the mass term in \eqnref{eq:MassContrGen} is independent of the gauge fields, the kinetic contribution has to be rephrased.
The kinetic contribution in \eqnref{eq:KineticContrGen} corresponds to the creation or annihilation of a particle-antiparticle pair on neighbouring lattice sites and the simultaneous adjustment of the electric field on the link in between. 
The green lines in \figref{fig:theory}(b) mark the fields $\hat{E}_{(0,0),\vec{e}_{x}}$, $\hat{E}_{(0,0),\vec{e}_{y}}$ and $\hat{E}_{(1,0),\vec{e}_{y}}$ which are automatically adjusted when charges are created.
This fact follows from our arbitrary choice of enforcing the three Gauss' law constraints on exactly those links. 
For any other link, we require combinations of raising and lowering operators $\hat{P}_{j}$ and $\hat{P}_{\mu}$ ($j=1,2,3$ and $\mu = x,y$) such that  the specific link is adjusted, while all others remain unchanged. 
As an example, let us consider the generation of a particle in position $(1,1)$ and an antiparticle in $(1,0)$.
This choice implies either that the electric field $\hat{E}_{(1,0),\vec{e}_{y}}$ has to decrease [which is automatically adjusted through the creation of a charge string], or that the electric field $\hat{E}_{(1,1),\vec{e}_{y}}$ has to increase and hence the rotator $\hat{R}_{3}$ has to decrease. 
However, this action changes the electric fields $\hat{E}_{(1,1),\vec{e}_{x}}$, $\hat{E}_{(0,1),\vec{e}_{y}}$ and $\hat{E}_{(1,0),\vec{e}_{x}}$ as well. 
To remedy that, we lower the rotator $\hat{R}_{2}$, adjusting $\hat{E}_{(1,1),\vec{e}_{x}}$ and $\hat{E}_{(1,0),\vec{e}_{x}}$, and raise the string $\hat{R}_{y}$ to compensate for the change in $\hat{E}_{(0,1),\vec{e}_{y}}$. 
Following the same procedure, the rules for translating the kinetic Hamiltonian of \eqnref{eq:KineticContrGen} into the language of rotators and strings read
\begin{align}
\hat{\Psi}_{(0,0)}^\dagger \hat{U}_{(0,0),\vec{e}_{x}}^\dagger \hat{\Psi}_{(1,0)} &\to \hat{\Psi}_{(0,0)}^\dagger \hat{\Psi}_{(1,0)}, \cr
\hat{\Psi}_{(1,0)}^\dagger \hat{U}_{(1,0),\vec{e}_{x}}^\dagger \hat{\Psi}_{(0,0)} &\to \hat{\Psi}_{(1,0)}^\dagger \hat{P}_x^\dagger  \hat{\Psi}_{(0,0)},\cr
\hat{\Psi}_{(1,0)}^\dagger \hat{U}_{(1,0),\vec{e}_{y}}^\dagger  \hat{\Psi}_{(1,1)} &\to \hat{\Psi}_{(1,0)}^\dagger \hat{\Psi}_{(1,1)}, \cr
\hat{\Psi}_{(1,1)}^\dagger \hat{U}_{(1,1),\vec{e}_{y}}^\dagger \hat{\Psi}_{(1,0)} &\to \hat{\Psi}_{(1,1)}^\dagger \hat{P}_y^\dagger \hat{P}_{2} \hat{P}_{3} \hat{\Psi}_{(1,0)}, \cr
\hat{\Psi}_{(0,1)}^\dagger \hat{U}_{(0,1),\vec{e}_{x}}^\dagger  \hat{\Psi}_{(1,1)} &\to \hat{\Psi}_{(0,1)}^\dagger \hat{P}_1 \hat{\Psi}_{(1,1)}, \cr
\hat{\Psi}_{(1,1)}^\dagger \hat{U}_{(1,1),\vec{e}_{x}}^\dagger \hat{\Psi}_{(0,1)}  &\to \hat{\Psi}_{(1,1)}^\dagger  \hat{P}_x^\dagger \hat{P}_{2} \hat{\Psi}_{(0,1)}, \cr
\hat{\Psi}_{(0,0)}^\dagger \hat{U}_{(0,0),\vec{e}_{y}}^\dagger \hat{\Psi}_{(0,1)}  &\to \hat{\Psi}_{(0,0)}^\dagger \hat{\Psi}_{(0,1)}, \cr
\hat{\Psi}_{(0,1)}^\dagger \hat{U}_{(0,1),\vec{e}_{y}}^\dagger \hat{\Psi}_{(0,0)}  &\to \hat{\Psi}_{(0,1)}^\dagger \hat{P}_y^\dagger \hat{\Psi}_{(0,0)} .
\end{align}
Inserting these into \eqnref{eq:KineticContrGen}, we obtain the kinetic contribution to the total Hamiltonian as
\begin{eqnarray}\label{eq:KineticContrSingle}
\hat{H}_K &=& \kappa \left[ \hat{\Psi}_{(0,0)}^\dagger (\mathbbm{1} + \hat{P}_x) \hat{\Psi}_{(1,0)} + \right. \nonumber \\
 && \left. \hat{\Psi}_{(0,1)}^\dagger (\hat{P}_1 + \hat{P}_{2}^\dagger \hat{P}_x) \hat{\Psi}_{(1,1)} + \right. \nonumber \\
  &&
\left. \hat{\Psi}_{(0,0)}^\dagger (\mathbbm{1} + \hat{P}_y) \hat{\Psi}_{(0,1)}+\right. \nonumber \\
&& \hat{\Psi}_{(1,0)}^\dagger ( \mathbbm{1} +
\left. \hat{P}_{2}^\dagger \hat{P}_{3}^\dagger \hat{P}_y) \hat{\Psi}_{(1,1)} + \t{H.c.} \right].
\end{eqnarray}

In conclusion, with the gauge part $\hat{H}_{\t{gauge}}$ of the Hamiltonian described by \eqnsref{eq:MagneticContrSingle}{eq:ElectricContrSingle} and the matter part $\hat{H}_{\t{matter}}$ by \eqnsref{eq:MassContrGen}{eq:KineticContrSingle}, the system is fully characterised. 

The effective Hamiltonian we derive here for a periodic plaquette can be extended to a torus of arbitrary size [see \secref{sec:generalTorus}] or
to $d-$dimensional lattices. For the latter, one chooses operators $\hat{R}_i$ that describe the total electric field circulating around the $i$-th plaquette. Furthermore, one defines $d$ operators corresponding to loops that circulate around the whole lattice ($\hat{R}_x$ and $\hat{R}_y$ in the two-dimensional case here). The charge strings are eventually defined by arbitrary paths to each lattice point starting from the origin, as we show in \secref{sec:generalTorus} for $d=2$.

We will use the just derived Hamiltonian to compute the expectation value of the plaquette operator
\begin{align}\label{eq:BoxOpDef}
  \langle\Box\rangle &= -\frac{g^2a^2}{V} \bra{\Psi_0}\hat{H}_{B}\ket{\Psi_0},
 \end{align}
where $\ket{\Psi_0}$ is the ground state, and $V$ the number of plaquettes in the lattice, $V=4$ in this case. 
The expectation value of the operator $\Box$ is defined as a dimensionless number, which is bounded by $\pm 1$ and proportional to the magnetic energy.

\section{Transformation into the magnetic representation}
\label{sec:transformation}
In the following, we describe a scheme that allows switching from the so-called electric representation, where $\hat{H}_E$ is diagonal, to the magnetic one, where $\hat{H}_B$ is diagonal.
Our method is based on the replacement of the $U(1)$ gauge group with the group $\mathbb{Z}_{2L+1}$, and an accompanying transition from the compact formulation to a completely compact formulation, where both field degrees of freedom are treated as compact variables. 
While this procedure is general, we illustrate it for the minimal periodic system introduced in \secref{sec:single_plaq_ham} and consider generalisations in \secref{sec:Generalisations}.

Before presenting the scheme, we discuss the following observations about the considered Hamiltonian, that is now reduced to the sum of \eqnsref{eq:MagneticContrSingle}{eq:ElectricContrSingle}, while all charges $\hat{q}_{\vec{n}}$ in \eqnref{eq:ElectricContrSingle} are set to zero.
In particular, the lowering (raising) operators $\hat{P}_{x}^{(\dag)}$ and $\hat{P}_{y}^{(\dag)}$ acting on the strings are solely contained in the now absent kinetic Hamiltonian in \eqnref{eq:KineticContrSingle}. 
The total Hamiltonian thus commutes with $\hat{R}_{x}$ and $\hat{R}_{y}$, i.e. $[\hat{H}_{\t{gauge}},\hat{R}_{x}]=[\hat{H}_{\t{gauge}},\hat{R}_{y}]=0$, and as a consequence the strings become constants of motion. 
The dynamics induced by the pure-gauge Hamiltonian are thus restricted to different subspaces defined by $\hat{R}_{\mu}\ket{r_{\mu}} = r_{\mu}\ket{r_{\mu}}$, for $\mu = x,y$.
Starting in \secref{sec:performance}, we will be interested in a ground state property, therefore we restrict ourselves to the subspace where both strings are confined to the vacuum.
The effective Hamiltonian of this subspace can be readily obtained by setting $\hat{R}_{x} = \hat{R}_{y} = 0$ in \eqnsref{eq:MagneticContrSingle}{eq:ElectricContrSingle} which yields
\begin{align}\label{eq:FullHamSingleGauge}
	\hat{H}\E &= \hat{H}_E\E + \hat{H}_B\E, \nonumber \\
	\hat{H}_E\E &= 2g^2 \left[\hat{R}_1^2  +\hat{R}_{2}^2 + \hat{R}_{3}^2 - \hat{R}_{2}( \hat{R}_{1} + \hat{R}_{3})\right], \nonumber \\
	\hat{H}_B\E &= -\frac {1}{2g^2a^2} \left[ \hat{P}_1 + \hat{P}_{2} + \hat{P}_{3} + \hat{P}_1 \hat{P}_{2} \hat{P}_{3} +\t{H.c.}\right],
\end{align}
where we introduced the superscript $(\mathrm{e})$ to emphasise is the \emph{electric} representation.

Since the three rotators possess discrete but infinite spectra, any numerical approach for simulating the Hamiltonian in \eqnref{eq:FullHamSingleGauge} requires a truncation of the Hilbert space. 
In the following, $l$ denotes the cut-off value which is identified by
\begin{eqnarray}
	\hat{R}_j \ket{r_j} = r_j \ket{r_j} \;\forall\, r_j = -l,-l+1,\dots,l. 
	\label{eq:def_truncation}
\end{eqnarray}
Thus, the action of the truncated lowering operators is given as
\begin{eqnarray}
	\hat{P}_j \ket{r_j} = \begin{cases} \ket{r_j-1}, & \t{if}\, r_j > -l\\
		0, & \t{if}\, r_j = -l.
	\end{cases}
	\label{eq:def_raising_truncation}
\end{eqnarray}
Note that the total dimension of the Hilbert space is reduced to $(2l+1)^3$, which is still challenging to simulate even for relatively small values of $l$. 
In particular, calculations in the weak coupling regime suffer from this severe limitation and until now, no practical methods to solve this issue have been available.

Let us now introduce a formulation that allows for an efficient representation of the Hamiltonian's eigenstates in the weak coupling regime, where $g\ll1$. 
It is based on the exchange of the continuous $U(1)$ group with the discrete group $\mathbb{Z}_{2L+1}$, which provides a discrete basis for the vector potential operators $\hat{A}_{\vec{n},\vec{e}_\mu}$ and enables a direct transformation into this dual basis via a Fourier transform.
The approach is motivated by the key observation that, in the electric representation, the Hamiltonians of the continuous $U(1)$ group and the discrete $\mathbb{Z}_{2L+1}$ group after truncation ($l<L$) are equivalent. 
The group $\mathbb{Z}_{2L+1}$ consists  of $2L+1$ elements, thus the parameter $L$ indicates the size of the Hilbert space. 
In particular, the rotators $\hat{R}_{j}$ and lowering (raising) operators $\hat{P}_{j}^{(\dag)}$ ($j=1,2,3$) take the form
\begin{eqnarray}
	\hat{R}_j \ket{r_j} &=& r_j \ket{r_j} \;\forall \, r_j = -L,\dots,L \nonumber \\
	\hat{P}_j \ket{r_j} &=& \begin{cases} \ket{r_j-1}, & \t{if}\, r_j > -L\\
		\ket{L}, & \t{if}\, r_j = -L.
	\end{cases}
	\label{eq:def_raising_ZN}
\end{eqnarray}
The only difference between the truncated $U(1)$ group and untruncated $\mathbb{Z}_{2L+1}$ group is the cyclic property of the lowering (raising) operator, which transforms $\ket{L}$ into $\ket{-L}$ (and vice versa). 
However, after a truncation of $\mathbb{Z}_{2L+1}$ with $l<L$, this property is lost, meaning that \eqnsref{eq:def_raising_truncation}{eq:def_raising_ZN} correspond to each other and the two truncated groups become equivalent.

For now, consider the Hamiltonian which employs the complete $\mathbb{Z}_{2L+1}$ group.
Importantly, the relations in \eqnref{eq:def_raising_ZN} resort to a compact description of the electric field since the spectra of the rotators and strings are constrained to the compact interval $[-L,L]$. 
We now introduce the following replacement rules for these operators,
\begin{align}\label{eq:op_replacement}
	\hat{R} &\mapsto \sum_{\nu=1}^{2L} f_\nu^{s} \sin\left(\frac{2\pi \nu}{2L+1}\hat{R}\right), \nonumber \\
	\hat{R}^2 &\mapsto \sum_{\nu=1}^{2L} f_\nu^{c} \cos\left(\frac{2\pi \nu}{2L+1}\hat{R}\right) +\frac{L(L+1)}{3} \eins,
\end{align}
which reassemble Fourier series expansions. 
Crucially, this replacement is exact, i.e. there is no truncation of the Fourier series.
Employing the fact that the spectrum of $\hat{R}$ is discrete and takes integer values, the periodicity of the trigonometric functions can be exploited, which allows one to perform a summation over all coefficients where the sine (cosine) is equivalent. 
Hence, a finite number of $2L$ coefficients remain, which take the form
\begin{align} 
	f_\nu^s =& \frac{(-1)^{\nu+1}}{2\pi} \left[\psi_0\left(\frac{2L+1+\nu}{2(2L+1)}\right)-\psi_0\left(\frac{\nu}{2(2L+1)}\right)\right] \\
	f_\nu^c =& \frac{(-1)^\nu}{4\pi^2} \left[\psi_1\left(\frac{\nu}{2(2L+1)}\right)-\psi_1\left(\frac{2L+1+\nu}{2(2L+1)}\right)\right].
\end{align}
Here, $\psi_{k}(\bullet)$ is the $k$-th polygamma function. 
Let us further remark that these rules can be extended to higher powers in the variables $\hat{R}$ than considered in \eqref{eq:op_replacement}.

This replacement turns out to be convenient for the basis transformation explained below.
Introducing the convention $\ket{\vec{r}}=\ket{r_1}\ket{r_2}\ket{r_3}$ and recalling \eqnref{eq:def_raising_ZN}, the electric contribution from \eqnref{eq:FullHamSingleGauge} reads
\begin{eqnarray}
	\hat{H}_{E}\E &=& 2g^2 \sum_{\vec{r}=-\vec{L}}^{\vec{L}} \, \sum_{\nu=1}^{2L}\, \bigg\lbrace f^c_\nu \sum_{j=1}^3\cos\left(\frac{2\pi \nu}{2L+1} r_j\right) \nonumber \\
	&& -f_\nu^s  \sin\left(\frac{2\pi \nu}{2L+1} r_2\right) \sum_{\mu=1}^{2L}f_\mu^s\bigg[\sin\left(\frac{2\pi \mu}{2L+1} r_1\right) \nonumber \\
	&&+\sin\left(\frac{2\pi \mu}{2L+1} r_3\right)\bigg] \bigg\rbrace \ketbra{\vec{r}}{\vec{r}}.
	\label{eq:HE_in_E}
\end{eqnarray}
Note that we use the notation $\sum_{\vec{r}=-\vec{L}}^{\vec{L}}$ to indicate that the sum collects all combinations of $r_j$, where $r_j\in[-L, L]$,  $j=1,2,3$ and we neglected the constant energy shifts introduced by \eqnref{eq:op_replacement}.
The $\mathbb{Z}_{2L+1}$ magnetic Hamiltonian $\hat{H}_B\E$ can be obtained by substituting the cyclic $\hat{P}_j$ of \eqnref{eq:def_raising_ZN} into \eqnref{eq:FullHamSingleGauge}.

We are now in a position to perform the switch to the dual basis.
As shown in \appref{app:fourier}, for any $\gamma\in\mathbb{N}$, the discrete Fourier transform $\hat{\mathcal{F}}_{2L+1}$ diagonalises the lowering operators as
\begin{eqnarray}
	\hat{\mathcal{F}}_{2L+1} \hat{P}_j^\gamma \hat{\mathcal{F}}_{2L+1}^\dagger = \sum_{r_j=-L}^L \t{e}^{i\frac{2\pi}{2L+1}\gamma r_j} \ketbra{r_j}{r_j}.
\end{eqnarray}
Hence, by applying the discrete Fourier transform to the total Hamiltonian we diagonalise the magnetic contributions, while sacrificing the diagonal structure in the electric part, i.e.
\begin{eqnarray}\label{eq:HE_in_B}
	\hat{H}_{E}\B &=& g^2 \sum_{\nu=1}^{2L} \bigg\lbrace f_{\nu}^c \sum_{j=1}^3 \hat{P}_j^\nu + \frac{f_\nu^s}{2} \left[\hat{P}_2^\nu - (\hat{P}_2^\dagger)^\nu \right] \nonumber \\ &\times &\sum_{\mu=1}^{2L} f_\mu^s \left[\hat{P}_1^\mu + \hat{P}_3^\mu \right] \bigg \rbrace + \t{H.c.},
\end{eqnarray}
and
\begin{eqnarray}\label{eq:HB_in_B}
	\hat{H}_{B}\B &=&  -\frac{1}{g^2a^2}\sum_{\vec{r}=-\vec{L}}^{\vec{L}} \bigg[ \cos\left(\frac{2\pi r_1}{2L+1}\right) \nonumber \\
	&\:& + \cos\left(\frac{2\pi r_2}{2L+1}\right) +  \cos(\frac{2\pi r_3}{2L+1})\nonumber \\
	&\:& +  \cos\left(\frac{2\pi (r_1+r_2+r_3)}{2L+1}\right)\bigg] \ketbra{\vec{r}}{\vec{r}}.
\end{eqnarray}
Note that we introduced the superscript $(\mathrm{b})$, which refers to the \emph{magnetic} representation of the Hamiltonian. 
Using this representation, computations in the weak coupling regime $g\ll 1$ can be performed efficiently, as a truncation $l$ now chooses a cut-off for the magnetic field energy.
We emphasize that although we employed the rotator formulation of the Hamiltonian, the just presented procedure is likewise valid for the link formulation utilizing the electric field operators. Indeed, the replacement rules in \eqnref{eq:op_replacement} are then formulated in terms of $\hat{E}$ instead of $\hat{R}$ and inserted into the Hamiltonian in \eqnref{eq:HKS}. The corresponding magnetic representation is analogously obtained via an application of the Fourier transform.

The parameter $L$ now affects the accuracy of the simulation. 
In fact, while $L$ is completely irrelevant in the electric representation (truncated $U(1)$ and truncated $\mathbb{Z}_{2L+1}$ are equivalent), it strongly influences the results derived in the magnetic representation. 
While examining the relationship between $L$ and $l$ in more detail in \secref{sec:performance}, we qualitatively discuss our procedure to simulate the $U(1)$ group with the two representations of $\mathbb{Z}_{2L+1}$ in the following. 
To be more precise, for any $g$ we might always formulate a sequence of approximating representations for any quantum state of the system in the computational basis defined by $\ket{\vec{r}}$, i.e.,
\begin{eqnarray}
	\lvert\psi\E(g)\rangle &=&  \sum_{\vec{r}=-\vec{\infty}}^{\vec{\infty}} p_{U(1)}(g,\vec{r}) \ket{\vec{r}} \nonumber \\
	& \approx & \sum_{\vec{r}=-\vec{L}}^{\vec{L}} p_{\mathbb{Z}_{2L+1}}(g, \vec{r}) \ket{\vec{r}} \nonumber \\
	& \approx & \sum_{\vec{r}=-\vec{l}}^{\vec{l}} p\E(g, \vec{r}) \ket{\vec{r}}.
	\label{eq:stateExp}
\end{eqnarray}
Here, $p\E$ denotes the expansion coefficients in the electric representation, with the subscript indicating the group to which they are referring to (no subscript stands for the truncated $\mathbb{Z}_{2L+1}$). 
The first approximation in \eqnref{eq:stateExp} is due to the transition from $U(1)$ to the $\mathbb{Z}_{2L+1}$ group, while the second approximation represents the \textit{truncation} from $(2L+1)^3$ down to $(2l+1)^3$ states. 

The same scheme exists for the magnetic representation, where the weights $p\B(g,\vec{r}, L)$ are used for the state $\ket{\psi\B (g,L)}$. 
In this case, however, the choice of $L$ is important.
While the truncated electric representation directly corresponds to the truncated and compact $U(1)$ formulation, the completely compact formulation employed in the magnetic representation is crucially affected by the level of discretisation $L$.
This relation is examined further in \secref{sec:fidelity}, where we study the convergence of the two representations to $U(1)$ for intermediate values of the coupling $g$.
Hence, in the remainder of this manuscript we consider the completely compact formulation for the magnetic representation only and resort to the compact formulation for the electric representation, i.e. to \eqnref{eq:FullHamSingleGauge} for the case of pure gauge.

\begin{figure}[t]
	\centering
	\includegraphics[width=.9\columnwidth]{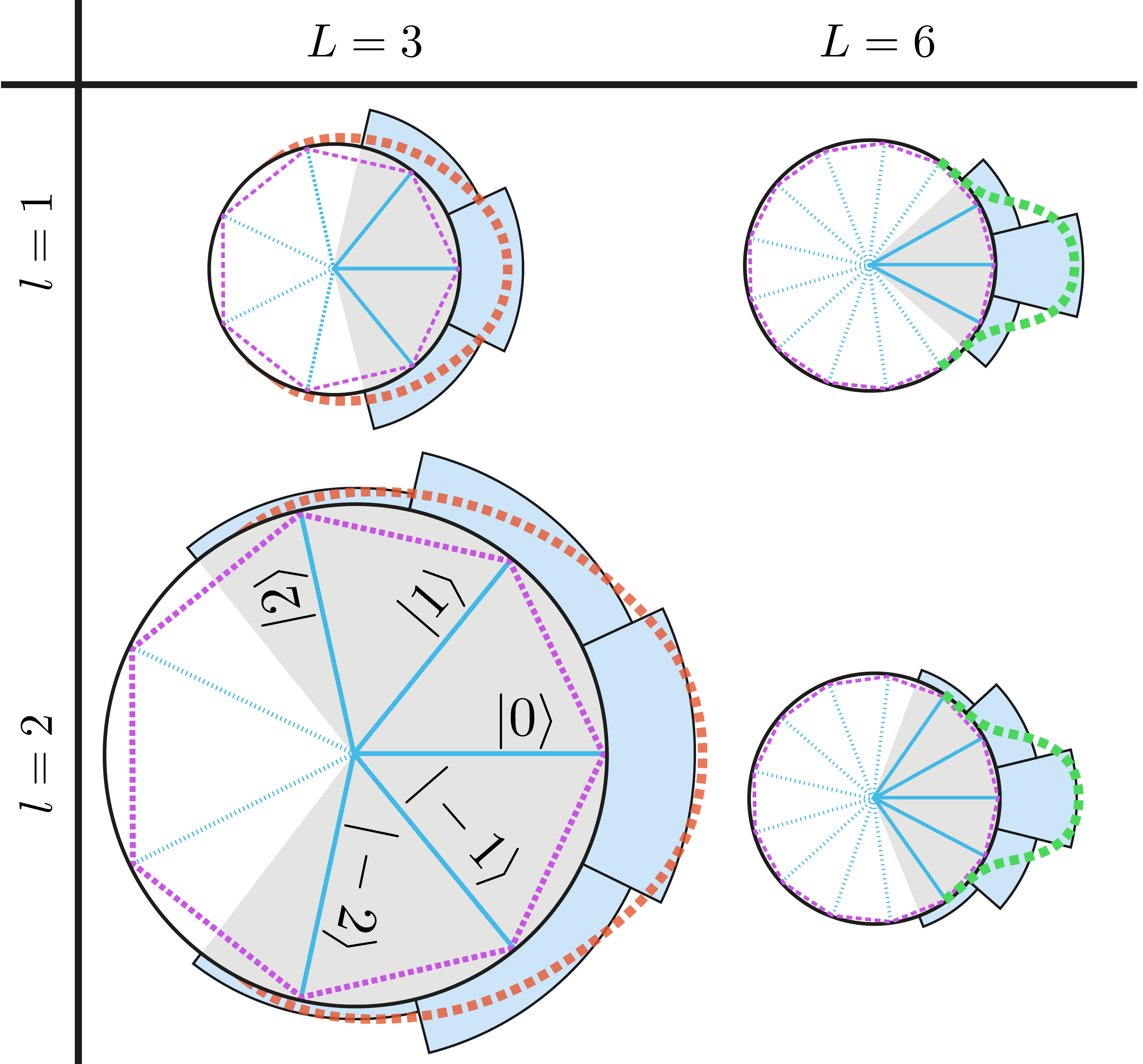}
	\caption{\textbf{Discrete approximation of a continuous distribution of states in the magnetic representation.} The ability to approximate a state is related to the quotient $l/L$. For a given $l$, $L$ controls the resolution of the approximation, which is always centred around the vacuum $\ket{\vec{0}}$. Black circles represent the $U(1)$ group, the violet $2L+1$ edged polygon the $\mathbb{Z}_{2L+1}$ group. Blue lines (solid and dashed) mark the $2L+1$ states of $\mathbb{Z}_{2L+1}$, while only the $2l+1$ states indicated with the solid lines are kept after truncating. Red and green markers are pictorial representations of states in $U(1)$ while the light blue areas correspond to their binned approximation.}
	\label{fig:U1_ZN_approximations}
\end{figure}

The interplay of the parameters $L$ and $l$ can be intuitively understood by employing a geometrical illustration. 
In \figref{fig:U1_ZN_approximations}, the black circles represent the continuous $U(1)$ group, which is approximated by $2L+1$ possible states (blue lines) of the $\mathbb{Z}_{2L+1}$ group.
For $l=L$, we faithfully describe the untruncated $\mathbb{Z}_{2L+1}$ group, and use both the solid and the dashed blue lines in the figure. 
By choosing $l < L$, we select the states marked with solid blue lines that lie symmetrically around $\ket{\vec{r}=\vec{0}}$ and achieve a binned approximation of any continuous $p_{U(1)}$ lying in the grey area.
Furthermore, for any fixed $l$, the parameter $L$ controls the spread of the available basis states (or bins) around $\ket{\vec{0}}$.
Since we are interested in the convergence of the truncated $\mathbb{Z}_{2L+1}$ to $U(1)$ which occurs for $L \to \infty$, we only consider the $2l+1$ states that are important for the dynamics. 
In particular, we disregard cyclic effects from the lowering operator $\hat{P}$ that are a distinctive feature of $\mathbb{Z}_{2L+1}$ with respect to $U(1)$ [see \eqnref{eq:def_raising_ZN}]. 

As an example of this relationship between $l$ and $L$, consider the two distributions drawn from $U(1)$, represented by the red and green dashed lines in \figref{fig:U1_ZN_approximations}. Clearly, the combination $L=3$, $l=1$ is insufficient to approximate the broad red distribution. 
Hence, we increase $l$ to completely cover the target distribution within the grey shaded area. 
A reduction of $L$ is also a practicable option, especially given the situation where $l$ could not be increased further because of a lack of computational resources.
By raising $L$ instead, our binned approximation has a higher resolution around the zero state. 
For the more localised green state, therefore, it is advantageous to choose the higher value $L=6$ when $l=2$ is an available option. 
In fact, the combination $L=6$, $l=1$ leads to a worse approximation of the green state than the choice $L=3$, $l=1$, which is therefore preferable when $l$ is limited to unity.

To that end, it is clear that the interplay of $L$ and $l$ represents a crucial point when estimating results and their error due to the performed discretisation.
Note that the spread of the distribution determines the value of the free parameter $L$, while $l$ will, for all practical purposes, be limited by the amount of physical resources, e.g. by the memory of a classical computer or the number of available qubits in a quantum simulator.

\section{Performance and application of the new approach}
\label{sec:performance}

In the previous section we outlined the transformation from the electric to the magnetic representation, suited to describe the strong and the weak coupling limits, respectively. 
Here, we develop a protocol which allows for assessing convergence of the truncated representations.

First, we qualitatively describe the system's behaviour at different values of the bare coupling, which will be useful for motivating the protocol.
Later, we consider the non-asymptotic cases, where it is not known whether a given truncation is sufficient to describe the considered system with the desired precision.
In our example dealing with a pure gauge $U(1)$ theory, this happens for $g \approx 1$, where both representations \emph{might} be inaccurate.  
Finally, we discuss convergence in the weak and strong coupling regimes, respectively and apply our protocol to estimate the plaquette expectation value. 
In particular, we consider the interplay between the parameters $l$ and $L$, which plays an important role when $g\gg 1$.

From now on, we work within a unit lattice spacing, i.e. $a=1$, but emphasise that this represents no restriction for the following results.

\subsection{Phenomenological analysis}\label{sec:phenomenology}

As mentioned above, the Hamiltonians in the electric and magnetic representations are related via the Fourier transform, i.e. the magnetic and electric fields are the dual of one another.
This relation consequently holds true for the eigenstates, illustrating the difficulty of expressing the ground state in the extremal regimes via the representation in which the dominant term of the total Hamiltonian is non-diagonal. 
For example, in the regime $g \ll 1$ the ground state is either determined by the bare vacuum $\ket{\vec{0}}$ or a superposition of all basis elements, depending whether the electric or magnetic representation is employed (both cases $\mathbb{Z}_{2L+1}$, $l=L$)
\begin{eqnarray}
|\t{GS}\B(g \ll 1,L)\rangle &=& \ket{\vec{0}} \nonumber\\ 
&=& \hat{\mathcal{F}}_{2L+1}^{-1}\sum_{\vec{r}} \frac{1}{(2L+1)^{3/2}} \ket{\vec{r}} \nonumber \\
&=& \hat{\mathcal{F}}_{2L+1}^{-1} |\t{GS}\E(g \ll 1)\rangle.
\label{eq:GS_magnetic_on_electric_side}
\end{eqnarray}
In other words, the coefficients $p(g \ll 1,\vec{r})=(2L+1)^{-3/2}$ in \eqnref{eq:stateExp} represent a uniform distribution assembling an equally weighted superposition of all basis states. 
This demonstrates the issue when truncating the Hilbert space by choosing $l<L$ and motivates the choice of switching to the magnetic representation. 
Note that in the limit $g \gg 1$ where the electric Hamiltonian is dominant, the roles of the two representations are interchanged.

While it is known that in the limit $g \rightarrow \infty$ ($g \rightarrow 0$) the ground state in the electric (magnetic) representation is the vacuum $\ket{\vec{0}}$, it is not clear what happens when $g$ deviates from these limits. 
In the following, we employ perturbation theory to estimate the required resources in order to describe the system. 
For any value of the bare coupling $g$, we determine the minimal truncation $l$ and resolution $L$ which suggest a high agreement with the untruncated and $U(1)$ theory.

Throughout the whole range of $g$ the system's ground state is center-symmetric, meaning that $p\E(\vec{r}) = p\E(-\vec{r})$ in \eqnref{eq:stateExp}.
This follows from the fact that the total Hamiltonian in \eqnref{eq:FullHamSingleGauge} is per-Hermitian \cite{Reid1997Some} (Hermitian with respect to the secondary diagonal; higher excited states can also be center-antisymmetric).
One can hence infer that the spread of the distribution $\lvert p\E(g,\vec{r}) \rvert$ in the electric representation is centred around $\ket{\vec{0}}$ and decreases with $g$, again motivating the developed basis transformation of the Hamiltonian. 
Equivalently, the same holds in the magnetic representation where the center-symmetric ground state becomes less localised by increasing $g$.

Employing the magnetic representation, we estimate the influence of the electric Hamiltonian with perturbation theory. 
For $g \rightarrow 0$, the unperturbed ground state is $| \t{GS}\B(g = 0,L)\rangle=\ket{\vec{0}}$, while the first order correction $|\t{GS}\B_{\t{corr}}\rangle$ takes the form
\begin{eqnarray}\label{eq:GScorrHB}
&&\lvert\t{GS}\B_{\t{corr}}(g, L)\rangle = \sum_{\substack{\vec{r}=-\vec{l},\\ \vec{r}\not=\vec{0}}}^{\vec{l}} \frac{\bra{\vec{r}}\hat{H}_E^{(b)}\ket{\vec{0}}}{E_{\vec{0}}-\bra{\vec{r}}\hat{H}_B^{(b)}\ket{\vec{r}}} \ket{\vec{r}}, \nonumber \\
&&\t{where} \;\;\; \left\lvert \frac{\bra{\vec{r}}\hat{H}_E^{(b)}\ket{\vec{0}}}{E_{\vec{0}}-\bra{\vec{r}}\hat{H}_B^{(b)}\ket{\vec{r}}} \right\rvert \lesssim  \frac{g^{4} \left(2L+1\right)^4}{|\vec{r}|^4}.
\end{eqnarray}
Here, we require $L\gg 1$ for the inequality, and we introduced the unperturbed ground state energy $E_{\vec{0}} = -4/g^2$.
Note that the chosen truncation $l$ determines the maximal length of $\vec{r}$ as $\lvert \vec{l} \rvert = \sqrt{3} l$, while the population in the states $\ket{\vec{r}}$ is proportional to $(gL/|\vec{r}|)^8$.
The upper bound on the population in each $\ket{\vec{r}}$ allows one to determine the part $p_{r>l}$ of the population that is left is out by the truncation at $l$, which yields $p_{r>l} \propto g^8 L^8/l^5$.
Hence, in order to cover the whole distribution by our truncation, we require $l^5 > g^8L^8$ such that large $\lvert \vec{r} \rvert$ states which are not covered by the truncation are only marginally populated.
Respectively, if the truncation $l$ is fixed, we infer that a resolution change $L\propto g^{-1}$ is required.
In fact, if $g^8L^8/|\vec{r}|^8$ takes large values for all $\vec{r}$, the chosen discretisation is not able to capture the spread of the true distribution, i.e. one would encounter the situation illustrated in the first row of \figref{fig:U1_ZN_approximations}. 

This can be intuitively understood by observing that the transition amplitudes between the ground state and states $\ket{\vec{r}}$ induced by $\hat{H}_E\B$ are suppressed by the respective energy gap, i.e. the denominator in \eqnref{eq:GScorrHB}.
The gap itself is controlled by $L$, which is a direct consequence of \eqnref{eq:HB_in_B}.

The analogous calculation for the electric representation yields $l > g^{-2} /\sqrt{3} $ which is independent of $L$. 
Here, the energy gap is not affected by $L$ and hence deviations from the continuum result have to be associated with a truncation $l$ which is insufficient for the state one aims to approximate.

Concluding, decreasing (increasing) $g$ requires additional computational resources in the electric (magnetic) representation.

\subsection{Fidelity and convergence of the two representations}\label{sec:fidelity}

This section is devoted to a convergence analysis, which examines the agreement between the two representations. 
Although we have developed a scheme that allows one to represent, discretise and truncate the Hamiltonian in the weak coupling regime, the optimal choice of the parameter $L$ is not clear a priori.
Clearly, $l$ should usually be chosen according to the availability of the computational resources, which then determines the most suitable $L$ depending on the bare coupling. 
Furthermore, there is an uncertainty regarding which representation to choose if one is not explicitly considering one of the extremal regimes, $g\gg1$ and $g\ll 1$. \\

We first develop a criteria to estimate the agreement of the two representations. 
Therefore, we employ their relation via a unitary transformation and define the \emph{Fourier fidelity} $F_{\rm f}$ with respect to the same state derived in both representations, e.g. an eigenstates belonging to the same eigenvalue of some observable, such as the ground state derived in both (truncated) representations for a fixed value of $g$. 
We write $F_{\rm f}$ as
\begin{equation}\label{eq:FidelityRepresentation}
F_{\rm f}(l) = \max_{L>l}\left| \bra{\psi\B(L,l)}\hat{\mathcal{F}}(L,l) \ket{\psi\E(l)} \right|^2,
\end{equation}
where the Fourier transform
\begin{equation}\label{eq:FidelityReps}
\hat{\mathcal{F}}(L,l) = \left(\frac{1}{\sqrt{2L+1}}\sum_{k,j = -l}^{l} \t{e}^{i\frac{2\pi}{2L+1}jk}\ketbra{j}{k}\right)^{\otimes 3}
\end{equation}
is truncated, i.e. the indices of the sums are limited by $\pm l$ instead of $\pm L$\footnote{Note that this is a consequence of the truncated Hilbert space. However, the operator in \eqref{eq:FidelityReps} not being unitary is no limitation, since the states in \eqnref{eq:FidelityRepresentation} could be embedded in the Hilbert space required by the full Fourier transform. Here, the coefficients for each basis element $c_n$ with $n>l$ are set to zero in both representations.}.
Due to the truncation, the features captured in both states are not necessarily equivalent which results in low values of the Fourier fidelity.
Vice versa, high values indicate that -- for the considered state -- the representations are equivalent and yield the same result. 
Note that this further suggest that the result is close to the hypothetical one derived within the untruncated theory, since the unification of both representations nearly covers the total Hilbert space\footnote{Recall that under the Fourier transform, local features are transformed into global ones and vice versa, e.g. a Gaussian is transformed into a Gaussian with inverse width.}.

Clearly, a low Fourier fidelity is not the only decisive criteria whether a derived result is robust against changes in $l$ or $L$, especially in the extremal regimes of the bare coupling where the truncation effects of the non-appropriate representation are severe. 
We thus employ the so called \emph{sequence Fidelity} $F_{\rm s}$, which measures the overlap of the same state (in the chosen representation) derived within successive values of truncations $l-1$ and $l$,
\begin{equation}\label{eq:s_fid}
F_{\rm s}^{(\mu)}(l,L) = \sum_{\vec{r}=-\vec{l}+\vec{1}}^{\vec{l-1}} \braket{\psi^{(\mu)}(l-1,L)}{\vec{r}} \braket{\vec{r}}{\psi^{(\mu)}(l,L)}.
\end{equation}
Here, $\mu = e,b$ indicates considered representation while $L$ is only present in the magnetic case (in the electric representation we can use the truncated $U(1)$ model). 
Since the truncated models converge to the untruncated $U(1)$ model in the limit $l \rightarrow \infty$, high values of $F_{\rm s}$ indicate, under a suitable assumption, that the chosen truncation $l$ is able to capture the whole distribution of the wave vector (as for the case $l=2$, $L=3$ in \figref{fig:U1_ZN_approximations}).
Such a conclusion can not be drawn in the case where the distribution is multimodal with disjoint fractions that lay outside the covered space. 
Then, the sequence fidelity yields high values for subsequent values of $l$ but would not for larger differences of the considered truncation.
Nevertheless, this represents a common issue present in all approaches employing truncation techniques that lack the exact true solution.\\

Let us now return to the ground state of the pure gauge model.
Due to their diagonal forms the electric (magnetic) representation yields more accurate results in the strong (weak) coupling regime, however there is no intuition for the intermediate regime $g \approx 1$. 
We hence calculate the Fourier fidelity to obtain an indicator whether results obtained via the different representations at finite values of $l$ are compatible, that is, whether the chosen truncation is enough to capture the local and non-local properties of the state vector. 
\begin{figure}[t]
	\centering
	\includegraphics[width=.99\columnwidth]{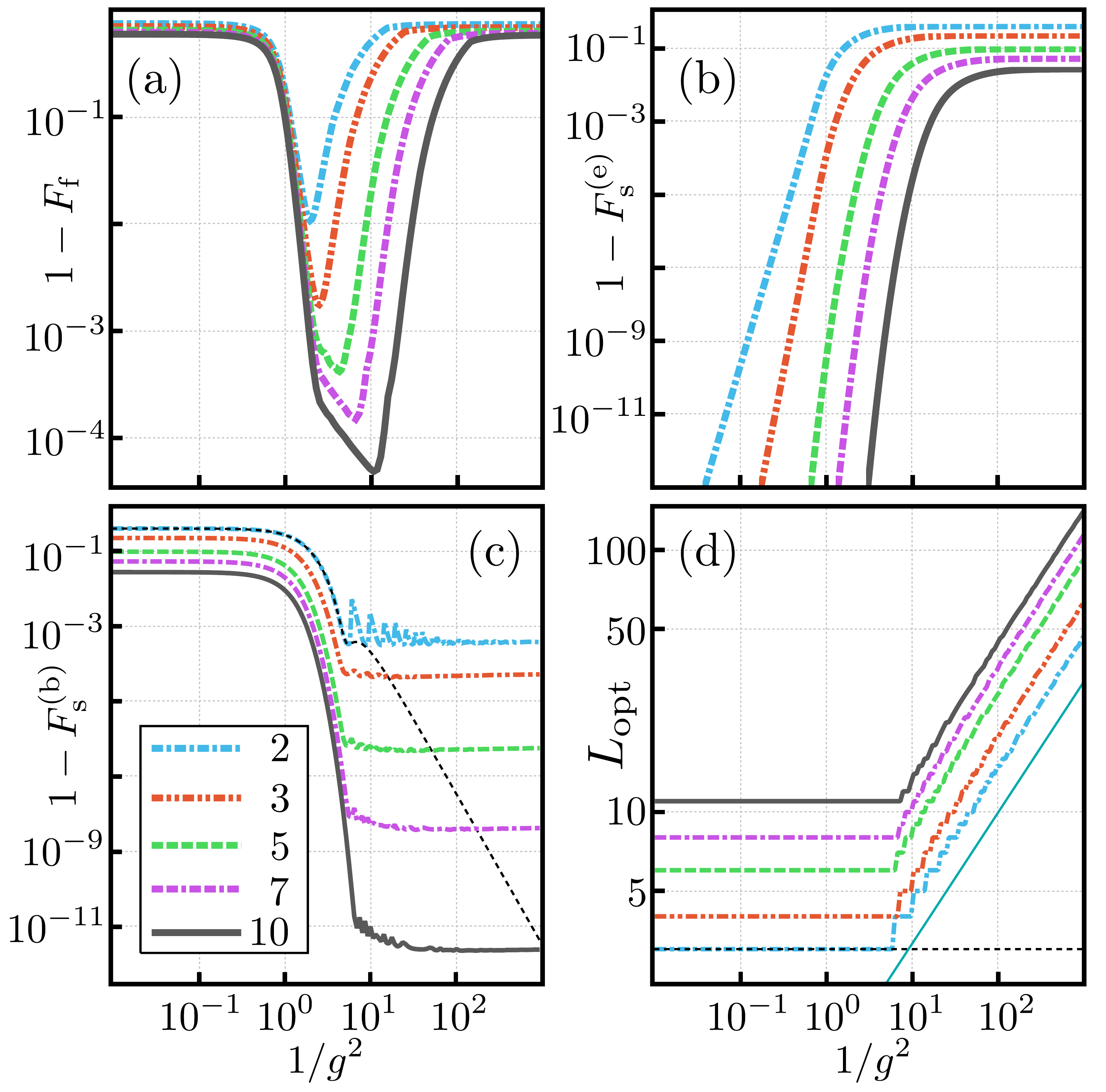}
	\caption{\textbf{Convergence analysis of the basis representations.} In (a), the Fourier infidelity in the intermediate region is decreasing with $l$ as the whole wave function can be captured by the truncation. The sequence infidelities in (b) and (c) illustrate convergence to the $U(1)$ theory and the freezing effect respectively. The values of $L$ optimizing the sequence fidelities of (c) are displayed in (d). Here, freezing is detected by curves similar to the black dashed lines.}
	\label{fig:state_convergence}
\end{figure}
\figref{fig:state_convergence}(a) illustrates the Fourier infidelity $1-F_{\rm f}(l)$ of the ground state for different values of $g^{-2}$. 
The global maximum of $F_{\rm f}$ arises in from the compromise of having truncation $l$ and resolution $L$ big enough to both contain and resolve the details of the state's distribution. 
For example, in \figref{fig:U1_ZN_approximations}, it becomes clear that an increase in resolution $L$ reduces the available domain to accommodate a distribution with too high spread if $l$ is not increased accordingly.
This relation is the origin of the kink in the Fourier fidelity appearing for larger $l$ in \figref{fig:state_convergence}(a), from where the decrease in the Fourier fidelity is solely attributed to an increase of the resolution.
Note that for $l=10$ we exceed a fidelity of $99.99\%$. \\

In the remainder of this section, we will focus on the strong and weak coupling regime where the Fourier fidelity is no meaningful quantity due to the inability to express the state within a truncated basis in both representations.
For the electric representation, the sequence Fidelity has a simple interpretation ($L$ is absent here) as it quantifies the overlap between the ground state obtained within different truncations. 
Since the energy spectrum is fixed and does not depend upon $L$, a unit value of $F_{\rm s}\E(l)$ implies that the considered state is unaffected by an increase in $l$. 
This suggests that higher truncations do not improve the result and that the model converged to the untruncated $U(1)$ ground state, which can be further motivated by examining the behaviour of $1-F_{\rm s}\E(l)$ in \figref{fig:state_convergence}(b).
As expected, in the strong coupling regime the sequence Fidelity approaches unity, indicating convergence to the untruncated model, where it is helpful to recall that the ground state in this limit is given by a single basis state, $\ket{\vec{0}}$. 
Approaching the intermediate regime $g \approx 1$, $F_{\rm s}\E(l)$ reduces to a $l$-dependent constant value, which indicates that the truncation is insufficient to describe all features of the ground state appropriately. \\

In the magnetic representation, the situation is substantially more complicated, since the approximation of the continuous $U(1)$ group with the discrete $\mathbb{Z}_{2L+1}$ group introduces the intricate interplay of $l$ and $L$. 
As mentioned above, higher values of $L$ allow for a better local approximation of the state which comes at the expense of the tails, which are cut off if $l$ is too small (see \figref{fig:U1_ZN_approximations}). 
In terms of the sequence fidelity $F_{\rm s}\B(l,L)$, this implies that for each value of $l$ there exists a unique optimal value $L_{\rm opt}$ of $L$. 
Let us stress that this is only true for the ground state of the pure gauge theory considered here. 
In a more general setting, possibly including matter and higher excited states, $F_{\rm s}\B(l,L)$ might have multiple optimal values of $L$.

Another complication is given by the fact that $L_{\rm opt}$ does not necessarily corresponds to the global maximum of the sequence fidelity. 
In particular, a \emph{freezing} effect can occur for highly localised distributions, where the resolution $L$ is insufficient to capture any of its features. 
Consequently $\ket{\psi\B(l,L)}$ and $\ket{\psi\B(l+1,L)}$ are practically the same state and thus yield high values of the sequence fidelity in \eqnref{eq:s_fid}. 
In the scenario examined here, the freezing mechanism can be observed in the weak coupling regime, where the ground state is highly localised around $\ket{\vec{0}}$. 
If $L$ is too small, i.e. the bin belonging to the latter state is to wide, all population is accumulated there and the state does not change if $g$ is decreased while $L$ is kept constant.
However, it is possible to identify the freezing effect by an educated interpretation of $F\B_{\rm s}$. 

\figref{fig:state_convergence}(c) illustrates that the sequence infidelity $1-F\B_{\rm s}(l,L_{\rm opt})$ in both regimes saturates at a $l$-dependent value. 
Analogous to the electric representation, it saturates in the strong coupling regime ($g\gg1$), however the saturation for weak coupling stems from the limited ability to approximate a continuous approximation with a fixed number of discrete levels. 
To be more precise, for every $l$ the optimal $L_{\rm opt}$ is chosen as the best compromise of resolution around $\ket{\vec{0}}$ and a proper representation of the tails of the distribution.
In \secref{sec:phenomenology}, we demonstrated that $L_{\rm opt}$ increases as $g$ is decreased, which we now confirm numerically in \figref{fig:state_convergence}(d) (see \appref{app:optimalL} for more details). 
Note that as soon as $L$ increases, it does so as $L\sim g^{-1}$, supporting the perturbative results before.
Physically speaking, $L$ increases with $g^{-1}$ since the spread of the population distribution in the magnetic representation decreases, and thus more resolution nearby the state $\ket{\vec{0}}$ is required. 

The black dashed line in \figref{fig:state_convergence}(c) corresponds to the global maximum of $F\B_{\rm s}(l=1,L)$ for all $g^{-2}$. 
It does not saturate and vanishes in the limit $g^{-2} \rightarrow \infty$. 
Comparison with the black dashed line in \figref{fig:state_convergence}(d), which indicates that $L_{\rm opt}\equiv l+1$ reveals this as a characteristic of the mentioned freezing effect.

Concluding, both the Fourier and the sequence fidelities in \eqnsref{eq:FidelityRepresentation}{eq:s_fid} are two tools to assess the convergence of and agreement between the two representations. 
While the sequence fidelity must be applied in the extremal regimes, the Fourier fidelity yields a valuable quantification of the combined capabilities of the two representations for intermediate values of the bare coupling. 
\subsection{Estimation of \EPO}
\label{sec:performance_of_plaquette}

We now apply the tools developed in \secref{sec:fidelity} to calculate the expectation value $\langle\Box\rangle$ as defined in \eqnref{eq:BoxOpDef}. 
The value of $\langle\Box\rangle$ with respect to the system's ground state is an important quantity in LGTs, as it can be related to the running of the coupling \cite{Papero2}.

In the absence of dynamical matter, the total Hamiltonian solely consists of the two gauge field contributions. 
Therefore, we may determine a value  $g_{\rm m}$ separating the regimes where either of the respective representations is advantageous. 

Let $g_{\rm m}$ be the value of $g$ for which the Fourier fidelity in \eqnref{eq:FidelityRepresentation} is maximal with respect to the ground state, i.e. 
\begin{equation}
F_{g_{\rm m}}(l) =  \max_{\substack{L>l \\ g}}\left| \bra{ \t{GS}\B(L,l,g) }\hat{\mathcal{F}}(L,l) \ket{\t{GS}\E(l,g)} \right|.
\end{equation}

Since the electric (magnetic) representation shows exceeding performance in the strong (weak) coupling regime, we can assume that for a given truncation $l$, the best approximation is achieved by considering the electric representation in the range $g \in [g_{\rm m}, \infty)$ and the magnetic one for $g \in [0, g_{\rm m}]$ (compare also \secref{sec:fidelity} and \figref{fig:state_convergence}).

\begin{figure}[t]
\centering
\includegraphics[width=.99\columnwidth]{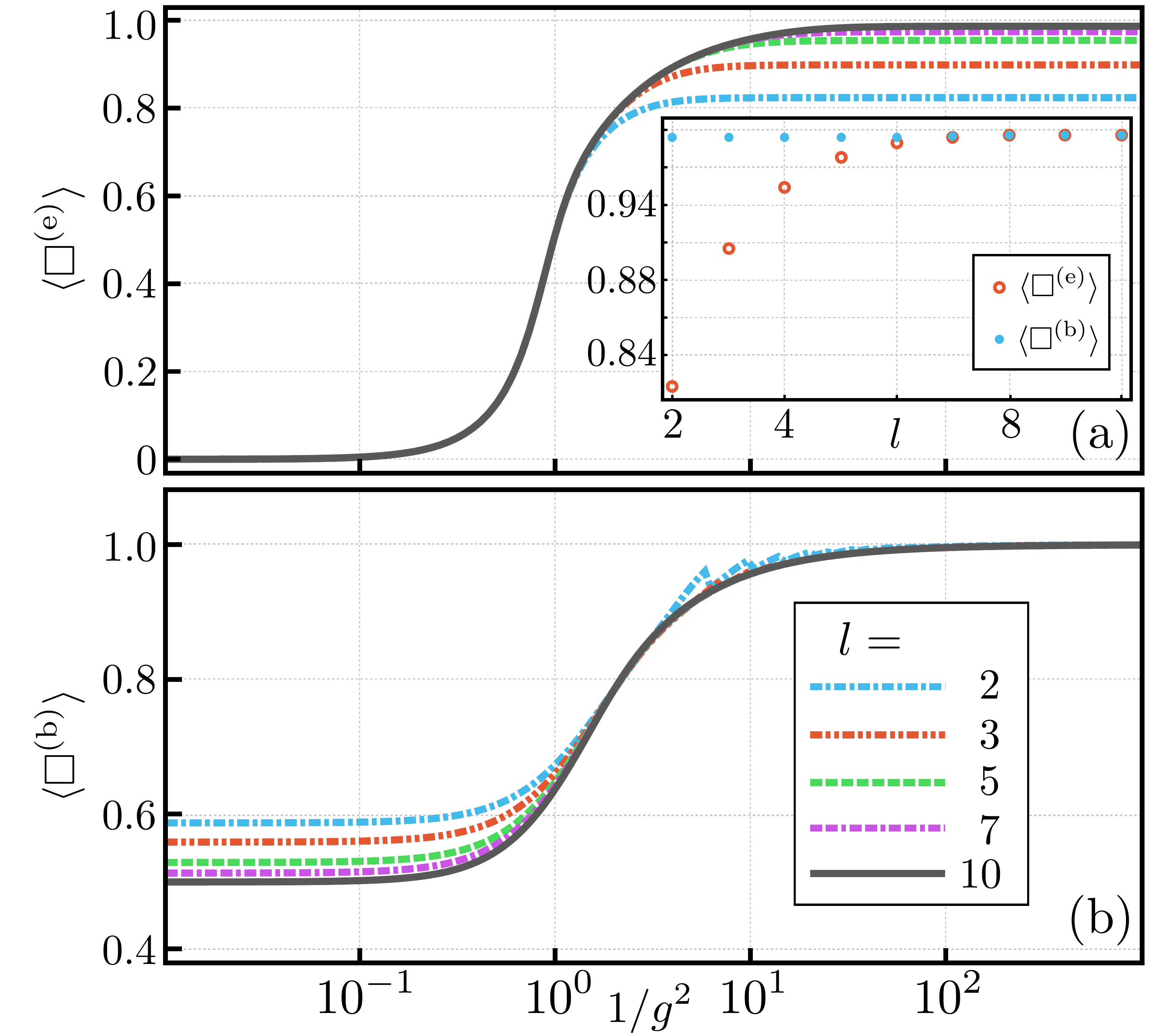}
\caption{\textbf{Estimation the plaquette operator.} Panel (a) displays the obtained curves in the electric representation, where the line styles correspond to different values of the truncation $l$. For the magnetic representation in panel (b), each point has been obtained via the optimisation of the sequence fidelity over $L$. We stress the considerably higher resource requirements ($l$) of the electric representation for calculations in the regime $g^{-2} > 1$. The inset in (a) shows the values for the different representations for all values of $l$ shown here when $g^{-2} = 10$.}
\label{fig:plaquette_expectation}
\end{figure}

\figref{fig:plaquette_expectation} shows  $\langle\Box\rangle$ for various truncations, derived both in the electric [panel (a)] and magnetic [panel (b)] representation. 
In the latter, we obtained the $L_{\rm{opt}}$ values that have been used for plotting via the sequence fidelity as described above.
From here, the true curve as it would be obtained from the untruncated $U(1)$ theory can be estimated via the asymptotic values of the different representations when the truncation $l$ is increased, since in the limit $l\rightarrow\infty$ both representations converge to the full theory.
We exemplify such an estimation with the inset in \figref{fig:plaquette_expectation}(a), that contains the results for different $l$ at $g^{-2} = 10$.
The convergence can be clearly observed, and both representations yield the same result up to the fourth decimal at $l=10$ ($\langle\Box\rangle = 0.9572 \pm 0.0001$).
Note that this convergence is not necessarily monotonic.
However, in the extremal regimes, we observe that the expectation value of $\Box$ increases with the truncation $l$ when employing the electric representation, while it decreases with the magnetic one, for which we will provide analytical arguments in \appref{app:asymptotic}.

To summarize this section, we recall that a naive approximation of $U(1)$ with $\mathbbm{Z}_{2L+1}$ (with $L$ fixed) leads to dramatically increasing computational costs when working on a wide range of $g$-values. 
As explained intuitively in \secref{sec:transformation}, the problem originates from the fact that $\mathbbm{Z}_{2L+1}$ converges not uniformly but pointwise to $U(1)$. 
For fixed resolution $L$ and fixed computational resources $l$, there is always a coupling $g$ small enough such that the $\mathbb{Z}_{2L+1}$ description displays freezing and hence cannot approximate the $U(1)$ continuum physics accurately. 
This can be understood by noting that the magnetic field Hamiltonian is gapless in both the continuum theory and in the $U(1)$-lattice description, but gapped in the  $\mathbbm{Z}_{2L+1}$-formulation. 
For fixed $L$ and decreasing $g$, the off-diagonal elements in the Hamiltonian $\hat{H}_E$ decrease with respect to the energy gap in $\hat{H}_B$ (as explained in more detail in \secref{sec:phenomenology}). 
If the energy in the system becomes comparable to the gap, the difference between $\mathbbm{Z}_{2L+1}$ and the true gauge group $U(1)$ becomes noticeable, which leads to the freezing effect (see \figref{fig:state_convergence}). 
Crucially, working with a value of $L$ suitable for the regime $g\ll 1$ will lead to exploding computational costs, i.e. will require very large values of $l$, in the intermediated coupling regime $g\approx 1$ to capture the relevant physics there.
Our solution to this problem is the dynamical adjustment of the parameter $L$ with the coupling $g$, that allows us to approximate $U(1)$ well for a wide range of couplings while including only a minimal number of states in our simulation (see \figref{fig:U1_ZN_approximations}).

\section{Generalisations: Dynamical matter and arbitrary torus}\label{sec:Generalisations}

In the following, we extend the results presented in \secref{sec:performance} by including staggered fermions in the numerical simulations. In particular, we show that matter does not introduce any fundamental complication for the completely compact formulation introduced in \secref{sec:2}. Moreover, to pave the way for further developments in the field, we derive the Hamiltonian for an arbitrary number of plaquettes on a torus with matter and periodic boundary conditions, and explain how to include static charges.

\subsection{Including dynamical charges}

Since the completely compact formulation only affects the gauge fields, the inclusion of matter is straightforward. 
Recall first the electric Hamiltonian in \eqnref{eq:ElectricContrSingle} and the substitution rules in \eqnref{eq:op_replacement}.
By using the relations for the Fourier transform derived in \appref{app:fourier}, the magnetic representation of the electric term in \eqnref{eq:ElectricContrSingle} is found to be
\begin{eqnarray}
\hat{H}_E\B &&= g^2 \sum_{\nu=1}^{2L}
\left\lbrace f_\nu^c \left[\K{1}+\K{2}+\K{3}+\frac{\K{x}+\K{y}}{2}\right] \right.\nonumber \\
&& + f_\nu^s \left[ \sum_{\mu=1}^{2L}f_\mu^s \left\lbrace\frac{1}{2}\hat{\mathcal{L}}_{2}^{\mu}\left(\L{1}+\L{3}\right)\left. \right. \right. \right. \nonumber \\ 
&& \left. \left. \left. -\frac{1}{4} \hat{\mathcal{L}}_{x}^{\mu}\left(\L{1}+\L{2}-\L{3}\right)+ \frac{1}{4} \hat{\mathcal{L}}_{y}^{\mu}\left(\L{1}-\L{2}-\L{3}\right) \right\rbrace \right. \right. \nonumber \\
&& \, \left. \left.  + i\frac{\hat{q}_{(1,0)}}{2} \left( \L{1}+\L{x}\right)\right. \right.\nonumber \\
&& \, \left. \left. + i\frac{\hat{q}_{(0,1)}}{2}  \left( \L{2}-\L{1}+\L{y} \right) \right. \right. \nonumber \\
&& \,  \left. \left. + i\frac{\hat{q}_{(1,1)}}{2} \left( 2\L{1}-\L{2}-\L{x} \right)\right] \right \rbrace \nonumber \\
&& \,  +g^2 \frac{\hat{q}_{(1,0)}^2 +  \hat{q}_{(0,1)}^2 + 2\hat{q}_{(1,1)}[\hat{q}_{(1,0)} +\hat{q}_{(1,1)}]}{2}.
\end{eqnarray}
For the sake of clarity, we defined the shorthand notations
\begin{eqnarray}
\K{j} = \hat{P}_j^\nu + (\hat{P}_j^\dagger)^\nu \;\; \t{and} \;\; \L{j} = \hat{P}_j^\nu - (\hat{P}_j^\dagger)^\nu.
\end{eqnarray}
The magnetic field Hamiltonian $\hat{H}_B\B$ remains the same as in \eqnref{eq:HB_in_B}, since it does not involve fermionic terms. 
However, the kinetic Hamiltonian in \eqnref{eq:KineticContrSingle} is modified in the presence of matter, yielding
\begin{eqnarray}\label{eq:KineticContrSingle_in_B}
\hat{H}_K\B &=& \kappa \sum_{\vec{r}=-\vec{L}}^{\vec{L}}  \left[ \hat{\Psi}_{(0,0)}^\dagger \left(1 + \t{e}^{-i\frac{2\pi}{2L+1}r_x}\right) \hat{\Psi}_{(1,0)} + \right. \nonumber \\
 && \left. \hat{\Psi}_{(0,1)}^\dagger \left(\t{e}^{-i\frac{2\pi}{2L+1}r_1} + \t{e}^{i\frac{2\pi}{2L+1}(r_2-r_x)} \right) \hat{\Psi}_{(1,1)} + \right. \nonumber \\
  &&
\left. \hat{\Psi}_{(0,0)}^\dagger \left(1 + \t{e}^{-i\frac{2\pi}{2L+1}r_y} \right) \hat{\Psi}_{(0,1)}+\right. \nonumber \\
&& \hat{\Psi}_{(1,0)}^\dagger \left( 1 +
\left. \t{e}^{i\frac{2\pi}{2L+1}(r_2+r_3-r_y)} \right) \hat{\Psi}_{(1,1)} \right. \nonumber \\ && \left.+ \mathrm{H.c.} \right.\Big] \ketbra{\vec{r}}{\vec{r}}.
\end{eqnarray}
In order to simulate fermionic matter, we recall the Jordan-Wigner transformation \cite{jordan1928pauli}
\begin{eqnarray}\label{eq:JWtrans}
\hat{\Psi}_{\vec{n}}^\dagger \mapsto  \prod_{\vec{l}<\vec{n}}(i\hat{\sigma}_z^{\vec{l}})\hat{\sigma}_-^n, 
\end{eqnarray}
where the vectorial relation $\vec{l} < \vec{n}$ is defined by $(0,0)<(0,1)<(1,1)<(1,0)$ to satisfy the fermionic commutation relations. 
In higher dimensions, it might however be useful to consider alternative approaches, such as fermionic projected entangled pair states or the elimination of the fermionic matter \cite{Kraus2010Fermionic, Zohar2018Eliminating}.
While we do not insert these equations into \eqnref{eq:KineticContrSingle_in_B}, we remark that the mass Hamiltonian in \eqnref{eq:MassContrGen} is simplified to
\begin{eqnarray}
\hat{H}_M = \frac{m}{2} \left(\hat{\sigma}_z^1-\hat{\sigma}_z^2+\hat{\sigma}_z^3-\hat{\sigma}_z^4\right),
\end{eqnarray}
which is independent of the chosen representation. 

Since these simulations are costly, i.e. the dimension of the truncated Hilbert space is given by $2^4(2l+1)^5$ (four charges, three rotators and two strings), we estimate the plaquette expectation value employing a harsh truncation of $l=2$, while fixing $L$ to the optimal values $L_{\rm opt}$ found in \secref{sec:fidelity} for the pure gauge case. This is a reasonable assumption, since strings and fermionic matter only play a role in the intermediate regime. Therefore, we can recover our previous results for the pure gauge theory in the strong and weak coupling limits, and focus our attention to the differences where $g \approx 1$.
We further introduce the mass and kinetic energy parameters as $m=\kappa=10$.

\begin{figure}[t]
\begin{center}
\resizebox{0.99\columnwidth}{!}{\includegraphics{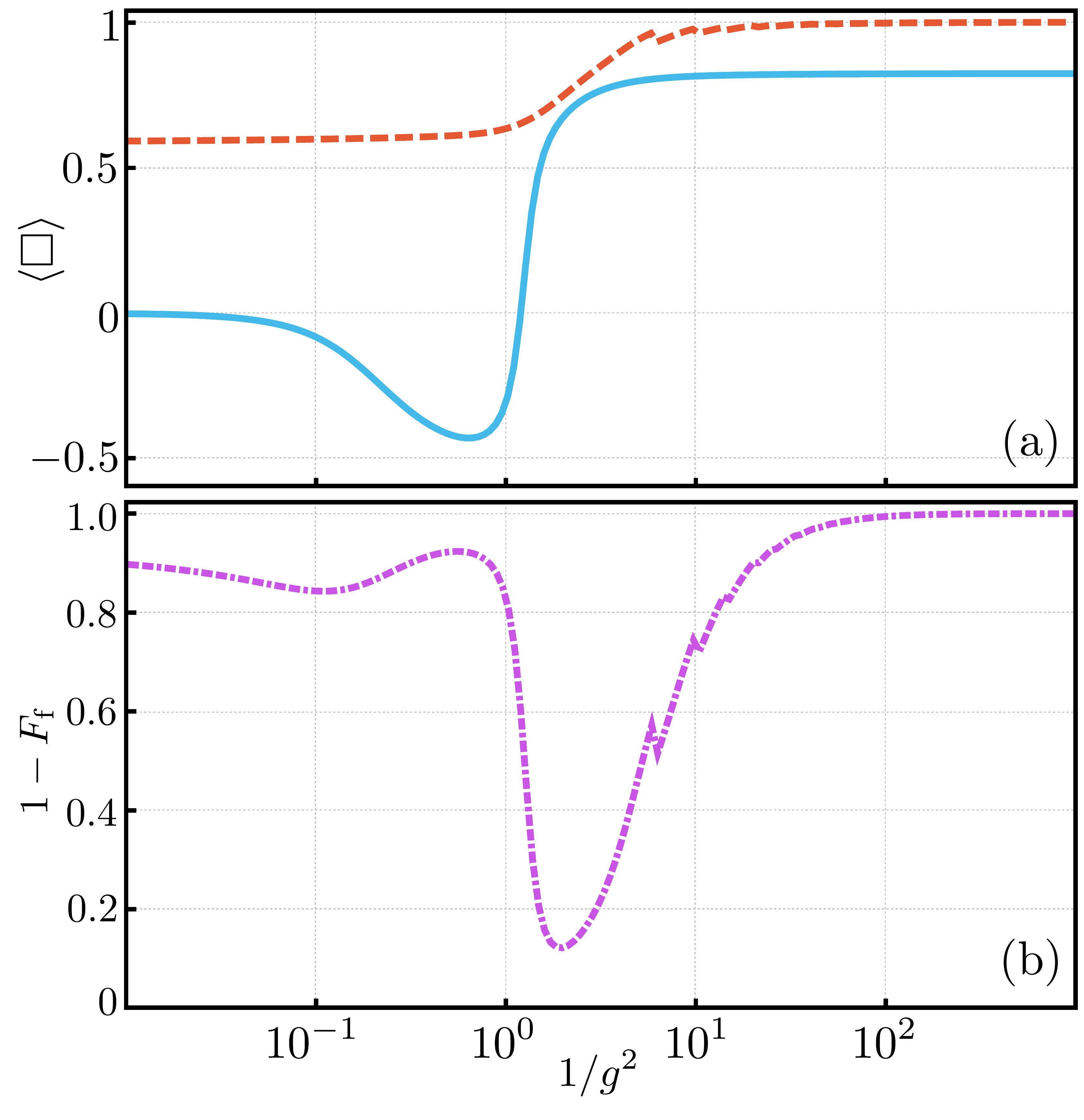}}
\caption{{\bf Plaquette expectation value in the presence of dynamical charges.} Panel (a) displays the expectation value for $l=2$ and (b) the Fourier fidelity derived in this case. The red dashed line in (a) corresponds to results derived in the magnetic representation, while the solid line is a result of the electric representation. For all curves we set $m=\kappa=10$.
}  
\label{fig:matterPlaquette}
\end{center}
\end{figure}

In \figref{fig:matterPlaquette}, we display the ground state expectation value $\langle\Box\rangle$ as a function of $g^{-2}$, together with the Fourier infidelity $1-F_{\rm f}(l)$. 
While the asymptotic regimes $g \ll 1$ and $g\gg 1$ show no qualitative difference if compared to the pure gauge case, the situation changes in the intermediate regime. 
There are novel features in both the electric and magnetic representations, such as the appearance of a negative dip. Nevertheless, we stress that conclusions drawn from this plot have to be taken with care, as the employed truncation limits the Fourier fidelity below $90\%$. 

Concluding, we demonstrated that our method is suitable to tackle simulations with matter, and can be scaled up to more complex systems.
A detailed analysis of novel effects and an accompanying study of the convergence is beyond the scope of this manuscript and left for future works.
\subsection{Hamiltonian for an arbitrary torus and charges}
\label{sec:generalTorus}
\begin{figure}[t]
\begin{center}
\resizebox{1.0\columnwidth}{!}{\includegraphics{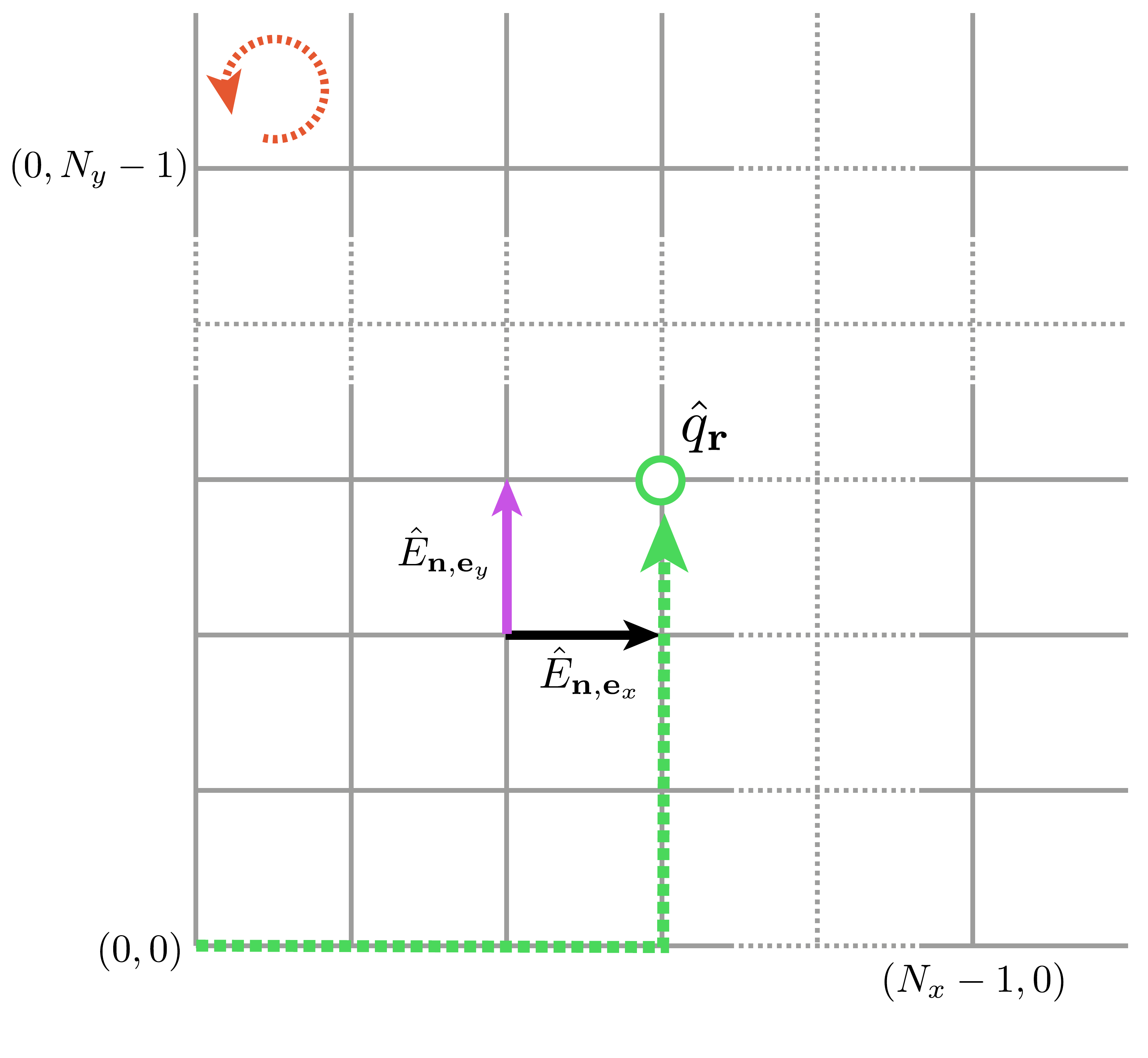}}\caption{\textbf{Periodic torus with charges.} We extend the construction of the periodic plaquette to a generic torus. We fix the rotator at $(0, N_y-1)$ to zero and choose the links' corrections to the electric field introduced by charges in accordance with the green dotted line. In particular, for any charge $\hat{q}_{\vec{r}}$, we connect the origin to the site $\vec{r}$ by moving first horizontally and then vertically.}  
\label{fig:generalization_torus}
\end{center}
\end{figure}

Here, we generalise the Hamiltonian of the minimal system considered in \secref{sec:single_plaq_ham} to any two-dimensional lattice with periodic boundary conditions.
As shown in \figref{fig:generalization_torus}, we extend the strategy used above to a torus of size $(N_x,N_y)$. By removing redundant DOF, we obtain the effective Hamiltonian in terms of two strings and $N_x N_y -1$ rotators. 
As before, we indicate each plaquette with its bottom-left site $\vec{n}=(n_x, n_y)$, where $n_x (n_y)= 0,\dots, N_x-1\, (N_y-1)$.
In addition, the rotator associated with the plaquette $\vec{n}$ is denoted by $\hat{R}_{\vec{n}}$, and the two strings by $\hat{R}_{x}$ and $\hat{R}_{y}$ (see \figref{fig:generalization_torus}).
This leads to $N_xN_y$ pairwise expressions for the electric fields,
\begin{eqnarray}
\hat{E}_{\vec{n},\vec{e}_x} &=& \delta_{n_y,0} \hat{R}_{x} + \hat{R}_{\vec{n}} - \hat{R}_{\vec{n}-\vec{e}_y} + \hat{q}_{\vec{n},x}, \nonumber \\  
\hat{E}_{\vec{n},\vec{e}_y} &=& \delta_{n_x,0} \hat{R}_{y} + \hat{R}_{\vec{n}-\vec{e}_x} - \hat{R}_{\vec{n}} + \hat{q}_{\vec{n},y}, 
\label{eq:ele}
\end{eqnarray}
where $\delta_{n,m} = 1$ for $n = m$, and zero otherwise. Moreover, $\hat{q}_{\vec{n},x}$ and $\hat{q}_{\vec{n},y}$ are the electric field's corrections due to the presence of dynamical charges, in accordance with Gauss' law. 
Since there are several ways to implement Gauss' law, a possible choice for $\hat{q}_{\vec{n},x}$ and $\hat{q}_{\vec{n},y}$ is (see the green lines in \figref{fig:generalization_torus})
\begin{eqnarray}
\hat{q}_{\vec{n},x} &=& -\sum_{r_x=n_x+1}^{N_x-1}\sum_{r_y=0}^{N_y-1} \delta_{n_y,0} \hat{q}_{\vec{r}}, \nonumber \\
\hat{q}_{\vec{n},y} &=& -\sum_{r_x=0}^{N_x-1} \sum_{r_y=n_y+1}^{N_y-1} \delta_{r_x,n_x} \hat{q}_{\vec{r}},
\label{eq:conventional_string}
\end{eqnarray}
where $\hat{q}_{\vec{n}}$ is the charge operator as defined in \eqnref{eq:charge_op}.
Note that also in this general case it is convenient to explicitly fix one of the rotators to zero, for instance $\hat{R}_{(0,N_y-1)} = 0$. 

Moving to the kinetic term, we employ the string convention presented in \eqnref{eq:conventional_string}, which yields the replacement rules (for details, see \appref{sec:detailsKinetic})
\begin{eqnarray}\label{eq:kin}
\hat{\Psi}_{\vec{n}}^\dagger && \hat{U}_{\vec{n},\vec{e}_x}^\dagger \hat{\Psi}_{\vec{n}+\vec{e}_x} \nonumber \\
&& \mapsto  \hat{\Psi}_{\vec{n}}^\dagger \left( \hat{P}_{x}^{-\delta_{n_x,N_x-1}}\,\prod_{r_y=0}^{n_y-1} \hat{P}_{(n_x,r_y)} \right) \hat{\Psi}_{\vec{n}+\vec{e}_x}, \nonumber \\
\hat{\Psi}_{\vec{n}}^\dagger && \hat{U}_{\vec{n},\vec{e}_y}^\dagger \hat{\Psi}_{\vec{n}+\vec{e}_y} \\
&& \mapsto  \hat{\Psi}_{\vec{n}}^\dagger 
\left(\hat{P}_{y}^{\dagger} \prod_{r_x=n_x}^{N_x-1} \prod_{r_y=0}^{N_y-1} \hat{P}_{(r_x,r_y)}\right)^{\delta_{n_y,N_y-1}} \hat{\Psi}_{\vec{n}+\vec{e}_y}. \nonumber 
\end{eqnarray}
From the above equations and from \eqnref{eq:ele}, we can calculate the components of the gauged Hamiltonian in the rotator and string basis as
\begin{eqnarray}\label{eq:gen_Ham}
\hat{H}_E &=&\frac{g^2}2 \sum_{\vec{n}} \hat{E}_{\vec{n},\vec{e}_x}^2 + \hat{E}_{\vec{n},\vec{e}_y}^2,  \nonumber \\
\hat{H}_B &=&- \frac{1}{2g^2a^2} \left(\prod_{\vec{n}\neq (0,N_y-1)} \hat{P}_{\vec{n}} + \sum_{\vec{n}\neq(0,N_y-1)} \hat{P}_{\vec{n}} \right) \nonumber \\
&+& \t{H.c.}, \nonumber \\
\hat{H}_K &=&\kappa \sum_{\vec{n}} \hat{\Psi}^\dagger_{\vec{n}} \left[\hat{\Psi}_{\vec{n}+\vec{e}_x}\hat{P}_{x}^{-\delta_{n_x,N_x-1}}\prod_{r_y=0}^{n_y-1} \hat{P}_{(n_x,r_y)} \right.\nonumber \\
&+& \left. \hat{\Psi}_{\vec{n}+\vec{e}_y} \left(\hat{P}_{y}^{\dagger} \prod_{r_x=n_x}^{N_x-1} \prod_{r_y=0}^{N_y-1} \hat{P}_{(r_x,r_y)}\right)^{\delta_{n_y,N_y-1}}\right] \nonumber \\
&+& \t{H.c.}, \nonumber \\
\hat{H}_M &=& m\sum_{\vec{n}} (-1)^{n_x+n_y}\hat{\Psi}_{\vec{n}}^\dagger \hat{\Psi}_{\vec{n}}.
\end{eqnarray} 
We note that the kinetic term contains string terms that depend on the choice of the background strings (see \figref{fig:generalization_torus}). Each of such terms can be simulated digitally with a number of gates that scales linearly with the system size.
We further remark that these equations are also valid for particles following bosonic statistics \cite{AMBJORN1989}, where the charge operator for site $\vec{n}$ is defined as
\begin{eqnarray}
\hat{q}_{\vec{n}} = q \, \hat{\Psi}_{\vec{n}}^\dagger \hat{\Psi}_{\vec{n}}.
\end{eqnarray}
In this case, the only required modification concerns the mass Hamiltonian, which becomes
\begin{eqnarray}
\hat{H}_M &=& m\sum_{\vec{n}} \,\hat{\Psi}_{\vec{n}}^\dagger \hat{\Psi}_{\vec{n}}.
\end{eqnarray}
Importantly, employing the relations derived in \appref{app:fourier} directly allows for the transformation between the electric and magnetic representations.

As a final remark, we highlight that it is possible to include static background charges into the description by keeping the $\hat{Q}_{\vec{n}}$ operators in \eqnref{eq:GaussKS}. These operators will then appear in the electric Hamiltonian, accompanying their corresponding dynamical charge $\hat{q}_{\vec{n}}$.

\section{Conclusions and outlook}
\label{sec:Conclusions}

We developed a new strategy for studying gauge theories. 
Our method is suited for simulations of fundamental particle interactions in all coupling regimes on near-term quantum computers (see \cite{Papero2}), as well as on classical devices. 
As a testbed, we applied our method to the lattice Hamiltonian formulation of QED in $2+1$ dimensions.

The key insight is the approximation of the continuous $U(1)$ gauge group with finite truncations of the $\mathbb{Z}_{2L+1}$ group, where $L \in \mathbb{N}$ can be arbitrarily large and is scaled with the value of the bare coupling $g$. 
This strategy allows us to work with fixed computational resources, i.e. including only a constant number of states in our simulation, for any value of $g$. 
At weak couplings we truncate the gauge fields in the magnetic representation of $\mathbb{Z}_{2L+1}$, while the truncation is performed in the electric representation for strong coupling. 

Depending on the regime, this solution offers to choose the representation with the smaller truncation error.
We benchmarked this novel regularisation scheme by computing the expectation value of the plaquette operator on a small periodic lattice, with and without dynamical matter, and estimated the accuracy of the computation.

Since our methods allows us to work at all values of $g$ and therefore at arbitrarily small values of the lattice spacing $a$, it provides the perspective to access, in principle, non-perturbative physics close to the continuum limit.
\newline
\\
\textbf{Quantum simulations:}\\

With regard to simulations of LGTs, there are two different lines of work. 
The first research line studies lattice models in their own right. Lattice gauge theories are for example relevant in condensed matter physics or can be interesting per se. 
The second line of research considers lattice gauge theories with the aim to study the underlying continuum theory which describes for example fundamental particle interactions and the standard model. 
For simulations of the latter, it is indeed of crucial importance that one is able to take the continuum limit of a lattice theory. In the field of quantum simulations, this challenge is mostly unanswered. 

So far, the only practical route to approach the continuum limit were analog quantum simulators using infinite degrees of freedom to represent the gauge fields. 
Neutral atoms in optical lattices offer a very good solution in this respect, as the gauge field can be represented as a spinor condensate while the charges are identified with moving fermions, and the gauge-matter interaction by spin changing collisions \cite{Zohar2012Simulating, Banerjee2012Atomic, Kasper2017Implementing, Dutta2017Toolbox, Barbiero2019Coupling, Kasper2020From}. 
A basic building block \cite{Mil2020A-scalable} and a one-dimensional proof-of-principle experiment \cite{Yang2020Observation} that exploit some of these ingredients have been already performed. 
Beyond one dimension, such approaches -- although very promising -- are fundamentally limited since magnetic (plaquette-type) interactions are realised via higher-order interactions. 
This results in low effective coupling strengths and thus in extremely challenging experimental requirements. 

While the analog approach based on bosonic degrees of freedom outlined above has the potential to achieve the continuum limit, the experimental realisation of two-dimensional theories involving magnetic terms is currently out of reach experimentally. 
This type of interaction is easier to realise digitally \cite{Tagliacozzo2013Optical, Tagliacozzo2013Simulation, Zohar2017Digital} and in qubit-based platforms \cite{Glaetzle2014Quantum, celi2019emerging}, such as trapped ions, Rydberg atoms and superconducting qubits. 
These simulation strategies however, currently lack the feature to reach the continuum limit.

Our new framework provides a route towards reaching the continuum limit in quantum simulators on different platforms and therefore opens a new perspective for meaningful simulations that address physical (i.e. continuum) phenomena that can be related to experiments in high energy physics. 
It will be interesting to explore the use of interacting chains of spins larger than spin-$1/2$ \cite{Senko2015Realization} to simulate gauge fields and to investigate the use of ultracold fermions, which will allow one to simulate fermions on a system that naturally displays the right quantum statistics. 
As a first step towards proof-of-concept demonstration using our new method, we show how to apply our scheme to current ion-based quantum hardware \cite{Papero2}. 
\newline
\\
\textbf{Tensor network calculations:}\\

Resource minimisation is especially relevant for classical simulations based on tensor network states. 
Our regularisation scheme can also be reinterpreted as a coupling dependent variational ansatz for the gauge-field states, where the ratio between the truncation parameter and the dimension of the discrete group $l/L$ is optimised. 
With this perspective in mind, our scheme could be combined with other variational approaches, e.g. with the method put forward in \cite{Bender2020Real-time}, with the aim to extend the continuum limit calculations beyond one dimension or for addressing toy models of high-energy physics, such as $CP(N-1)$ theories \cite{Laflamme2016CPN1}.
\newline
\\
\textbf{Extensions to higher-dimensional non-Abelian gauge groups:}\\

We note that both the minimal Hamiltonian formulation (in which redundant degrees of freedom have been removed) and our new regularisation scheme can be extended to higher dimensions and to non-Abelian gauge theories, also beyond the Hamiltonian approach. 
The solutions we proposed here are interesting for Lagrangian-based formalisms like the tensor approach \cite{Zhang2018Quantum, Meurice2020Discrete} or for the development of  novel Monte Carlo approaches to avoid the sign problem \cite{Ammon2016, GENZ2003187}. 
From a geometric perspective, once $U(1)$ is identified with $S^1$, and the magnetic vacuum with the north pole [see \figref{fig:U1_ZN_approximations}], our regularisation scheme at weak couplings can be interpreted as a lattice discretisation of a circle around the north pole. 
One could exploit the map of $SU(N)$ groups to higher dimensional spheres considered in \cite{Ammon2016,GENZ2003187} and repeat a similar procedure. 
An alternative discretisation of non-Abelian groups for classical and quantum simulations has also been considered in \cite{Hackett2019Digitizing, Alexandru2019Gluon, Lamm2019General, Lamm2020Parton}.

In conclusion, our work opens new perspectives for resource efficient Hamiltonian-based simulations and provides a concrete step in the `quantum way' to non-perturbative phenomena in high energy physics.

\section{Acknowledgements}

We thank Raymond Laflamme, Peter Zoller, and Rainer Blatt for fruitful and enlightening discussions.
This work has been supported by the Transformative Quantum Technologies Program (CFREF), NSERC and the New Frontiers in Research Fund.
JFH acknowledges the Alexander von Humboldt Foundation in the form of a Feodor Lynen Fellowship.
CM acknowledges the Alfred P. Sloan foundation for a Sloan Research Fellowship.
AC acknowledges support from the Universitat Aut\`{o}noma de Barcelona Talent Research program, 
from the Ministerio de Ciencia, Inovaci\'{o}n y Universidades  (Contract No. FIS2017-86530-P), 
from the the European Regional Development Fund (ERDF) within the ERDF Operational Program of Catalunya (project QUASICAT/QuantumCat),
and from the the European Union's Horizon 2020 research and innovation programme under the Grant Agreement No. 731473 (FWF QuantERA via QTFLAG I03769).
%
%
%
%
%
%
\bibliographystyle{apsrev4-1}
\bibliography{EfficientSim}

%
%

\appendix
\section{Dimensions of the (2+1) dimensional QED Hamiltonian}
\label{app:dimensions}
This appendix provides the dimensional analysis of the Hamiltonians used throughout this work. 
For the sake of simplicity and  to avoid problems when switching to the rotator formulation in \secref{sec:single_plaq_ham}, we aim for using dimensionless gauge field and matter operators.
Using natural units $c=\hbar=1$ results in the following relation of units,
\begin{eqnarray}
[\t{length}] = [\t{time}] = [\t{energy}]^{-1} = [\t{mass}]^{-1}.
\end{eqnarray}
The transition to the continuum limit is set by
\begin{eqnarray}
a^2 \sum_{\vec{n}} \rightarrow \int \mathrm{d}^2 x \quad (a \rightarrow 0),
\end{eqnarray}
where we denote the lattice spacing by $a$ and label all lattice points with a vector $\vec{n}$ as introduced in \secref{sec:QEDin2D}.
Since $a$ is a length, $[a] = \mathrm{mass}^{-1}$.
In the following, we discuss each part of the total Hamiltonian separately, keeping in mind that its dimension $[H] = [\t{energy}]$. 
\newline
\\
\textbf{Electric Hamiltonian}\\
The electric energy is given by
\begin{eqnarray}
	\hat{H}_E =  a^2 \sum_{\vec{n}}\frac{\tilde{g}^{2}}{2}\left(\hat{J}^{2}_{\vec{n},\vec{e}_x} + \hat{J}^{2}_{\vec{n},\vec{e}_y}\right), 
\end{eqnarray}
where the term in the sum has the units of a two-dimensional energy density, $[\t{energy}]/[\t{length}]^2=[\t{energy}]^3$. We now rescale the electric field operators $\hat{J}_{\vec{n},\hat{e}_\mu}$ as $a^3\hat{J}^2 = \hat{E}^2$ and absorb the remaining units into $g = \tilde{g} /\sqrt{a}$.
This yields $[g^2]=[\t{energy}]$ and 
\begin{eqnarray}
	\hat{H}_E = \frac{g^{2}}{2} \sum_{\vec{n}}\left(\hat{E}^{2}_{\vec{n},\vec{e}_x} + \hat{E}^{2}_{\vec{n},\vec{e}_y}\right),
\end{eqnarray}
where the field operators $\hat{E}_{\vec{n},\vec{e}_\mu}$ are dimensionless.
\newline
\\
\textbf{Magnetic Hamiltonian}\\
The plaquette operators are dimensionless since they are constructed from the field creation operators, which themselves are dimensionless.
Note that
\begin{eqnarray}
\hat{U} = \mathrm{e}^{iag\hat{A}},
\end{eqnarray}
where the product $ag\hat{A}$ needs to be dimensionless (allowing for a valid series representation).
Therefore, the dimension of the gauge potential is fixed by $[\hat{A}] = \sqrt{[\mathrm{mass}]}$. 
Recalling \eqnref{eq:PlaquetteOpDef}, the $\hat{P}_{\vec{n}}$ operator can be expressed as
\begin{eqnarray}
\hat{P}_{\vec{n}} = \t{exp} \left \lbrace ia^2g\left( \frac{\hat{A}_{\vec{n},\vec{e}_x}-\hat{A}_{\vec{n}+\vec{e}_y,\vec{e}_x}}{a} \right. \right. \nonumber \\
\left.\left.  -\frac{\hat{A}_{\vec{n},\vec{e}_y}-\hat{A}_{\vec{n}+\vec{e}_x,\vec{e}_y}}{a}\right) \right \rbrace.
\end{eqnarray}
Thus, in order to obtain the required continuum limit where $\hat{H}_B$ converges to $\int \t{d}x^2\, F_{\mu\nu}F^{\mu\nu}/4$, we define
\begin{eqnarray}
\hat{H}_B = -\frac{1}{2g^{2}} a^2\sum_{\vec{n}} \frac{\left(\hat{P}_{\vec{n}} + \hat{P}_{\vec{n}}^{\dag}\right)}{a^4},
\end{eqnarray}
where the the desired second order of $\hat{P}^\dagger + \hat{P}$ is proportional to $a^4$.
The denominator hence ensures its survival and thus the correct continuum limit.
This yields 
\begin{eqnarray}
\hat{H}_B = -\frac{1}{2g^{2}a^2} \sum_{\vec{n}} \left(\hat{P}_{\vec{n}} + \hat{P}_{\vec{n}}^{\dag}\right),
\end{eqnarray}
which is consistent with $[g^2] = [\mathrm{mass}]$ since we require that $[1/(g^2a^2)] = [\mathrm{mass}] = [\mathrm{energy}]$. 
\newline
\\
\textbf{Mass Hamiltonian} \\
We have that 
\begin{eqnarray}
\hat{H}_{M} &=& M a^2\sum_{\vec{n}} (-1)^{n_x+n_y} \hat{\psi}^\dag_{\vec{n}} \hat{\psi}_{\vec{n}}, 
\end{eqnarray}
where $\hat{\psi} = \hat{\phi}/a$ is the fermion field and $M$ represents the bare mass. 
Hence, the Hamiltonian reduces to
\begin{eqnarray}
\hat{H}_{M} &=& \frac{m}{a^2} a^2\sum_{\vec{n}} (-1)^{n_x+n_y} \hat{\phi}^\dag_{\vec{n}} \hat{\phi}_{\vec{n}} \nonumber \\
&=& m \sum_{\vec{n}} (-1)^{n_x+n_y} \hat{\phi}^\dag_{\vec{n}} \hat{\phi}_{\vec{n}},
\end{eqnarray}
where the operators $\hat{\phi}_{\vec{n}}$ are dimensionless and $[M]=\mathrm{mass}$.
\newline
\\
\textbf{Kinetic Hamiltonian}\\
Following the arguments for the mass and magnetic Hamiltonians, i.e. replacing $\hat{\psi} = \hat{\phi}/a$, we arrive at
\begin{eqnarray}
\hat{H}_{K} = K \sum_{\vec{n}} \sum_{\mu = x,y} (\hat{\phi}_{\vec{n}}^{\dag} \hat{U}_{\vec{n},\vec{e}_\mu} \hat{\phi}_{\vec{n}+\vec{e}_\mu} + \t{H.c.}),
\end{eqnarray}
where $K = 1/(2a)$. 
\newline
\\
\textbf{Rescaling of the fermionic field operators} \\
Note that it is possible to redefine $\hat{\phi} = \hat{\Psi}/\sqrt{\alpha}$, where $\alpha$ is dimensionless and $\hat{\Psi}$ is a rescaled fermion field.
By doing so,
\begin{eqnarray}
\hat{H}_{M} = \frac{M}{\alpha} \sum_{\vec{n}} (-1)^{n_x+n_y} \hat{\Psi}^\dag_{\vec{n}} \hat{\Psi}_{\vec{n}},
\end{eqnarray}
and
\begin{eqnarray}
\hat{H}_{K} = \frac{K}{\alpha} \sum_{\vec{n}} \sum_{\mu = x,y} (\hat{\Psi}_{\vec{n}}^{\dag} \hat{U}_{\vec{n},\vec{e}_\mu} \hat{\Psi}_{\vec{n}+\vec{e}_\mu} + \t{H.c.}),
\end{eqnarray}
where we might define the new effective mass $m = M/\alpha$ and effective kinetic energy scale $\kappa = K/\alpha$ valid for the rescaled fermionic fields as employed in the main text.
Moreover, this rescaling implies a rescaling of the charge operator, which we define as 
\begin{eqnarray}\label{eq:resc_charge_op}
\hat{q}_{\vec{n}} &=& Q ||\hat{\phi}_{\vec{n}}^\dagger\hat{\phi}_{\vec{n}}||_\t{max} \nonumber\\ 
&& \times \left( \frac{\hat{\phi}_{\vec{n}}^\dagger\hat{\phi}_{\vec{n}}}{||\hat{\phi}_{\vec{n}}^\dagger\hat{\phi}_{\vec{n}}||_\t{max}} - \frac{\eins}{2}[1-(-1)^{n_x+n_y}]\right) \nonumber \\
&=& q \left( \hat{\Psi}_{\vec{n}}^\dagger \hat{\Psi}_{\vec{n}} - \frac{\eins}{2}[1-(-1)^{n_x+n_y}]\right),
\end{eqnarray}
where $q = Q ||\hat{\Psi}_{\vec{n}}^\dagger\hat{\Psi}_{\vec{n}}||_\t{max} = Q/\alpha \in \mathbb{Z}$ and $q=1$ in the main  text. 
Note that this operator is dimensionless as well, since it is required to be of the same dimension as the electric field $\hat{E}$. 

\section{Hamiltonian in the link formalism and link-to-rotator translation rules}

\subsection{Effective Hamiltonian for a minimal lattice in the link formulation}
\label{app:Reparametrisation}
As explained in \secref{sec:2}, there are several ways to express the Hamiltonian of one or more plaquettes. 
In the main text, we employed electric loops for parametrizing the physical states and defining the corresponding operators, rotators and strings. These represent a natural choice since the rotator's lowering operators are directly identified with the plaquette operators $\hat{P}_{\vec{n}}$ [see \eqnref{eq:PlaquetteOpDef}]. 
Here, we derive the Hamiltonian of a lattice containing four sites and staggered fermions in terms of the links $\hat{E}_{\vec{n},\vec{e}_{\mu}}$.
In particular, we choose three out of the eight electric fields on the links, and express them in terms of the others to minimise the number of degrees of freedom. 
In this appendix, we consider the compact $U(1)$ formulation of the QED lattice Hamiltonian reviewed in \secref{sec:QEDin2D}.
The completely compact $\mathbb{Z}_{2L+1}$ QED formulation for both the electric and the magnetic representation can be obtained following the procedure outlined in \secref{sec:single_plaq_ham}.
Generalisations to multiple plaquettes are straightforward.

Employing \eqnref{eq:GaussKS}, we can directly assess Gauss' laws in \figref{fig:theory}(b) which yield
\begin{align}\label{eq:GaussLawSinglePlaquette}
\hat{E}_{(0,0),\vec{e}_{x}} + \hat{E}_{(0,0),\vec{e}_{y}} - \hat{E}_{(1,0),\vec{e}_{x}} - \hat{E}_{(0,1),\vec{e}_{y}} &= \hat{q}_{(0,0)} ,\nonumber \\  
\hat{E}_{(0,1),\vec{e}_{x}} + \hat{E}_{(0,1),\vec{e}_{y}} - \hat{E}_{(1,1),\vec{e}_{x}} - \hat{E}_{(0,0),\vec{e}_{y}} &= \hat{q}_{(0,1)} ,\nonumber \\  
\hat{E}_{(1,1),\vec{e}_{x}} + \hat{E}_{(1,1),\vec{e}_{y}} - \hat{E}_{(0,1),\vec{e}_{x}} - \hat{E}_{(1,0),\vec{e}_{y}} &= \hat{q}_{(1,1)} ,\nonumber \\  
\hat{E}_{(1,0),\vec{e}_{x}} + \hat{E}_{(1,0),\vec{e}_{y}} - \hat{E}_{(0,0),\vec{e}_{x}} - \hat{E}_{(1,1),\vec{e}_{y}} &= \hat{q}_{(1,0)}.
\end{align}
Only three of these constraints are independent since charge conservation requires $\hat{q}_{(0,0)} = - \hat{q}_{(0,1)} - \hat{q}_{(1,1)} - \hat{q}_{(1,0)}$. 
Expressing the arbitrarily chosen electric field operators $\hat{E}_{(1,0),\vec{e}_{x}}$, $\hat{E}_{(0,1),\vec{e}_{y}}$ and $\hat{E}_{(1,1),\vec{e}_{y}}$ in terms of the others, we write the constrained electric Hamiltonian as
\begin{eqnarray}
\hat{H}_E =&& \frac{g^2}{2} \left\lbrace \hat{E}_{(0,0),\vec{e}_{x}}^{2} + \hat{E}_{(0,0),\vec{e}_{y}}^{2} + \hat{E}_{(0,1),\vec{e}_{x}}^{2} + \hat{E}_{(1,0),\vec{e}_{y}}^{2}  \right.\nonumber \\
&& + \left.  \hat{E}_{(1,1),\vec{e}_{x}}^{2} + \left[ \hat{E}_{(0,0),\vec{e}_{y}} - \hat{E}_{(0,1),\vec{e}_{x}} + \hat{E}_{(1,1),\vec{e}_{x}} \right.\right.\nonumber \\
&& + \left.\left. \hat{q}_{(0,1)} \right]^{2} + \left[ \hat{E}_{(0,1),\vec{e}_{x}} - \hat{E}_{(1,1),\vec{e}_{x}} + \hat{E}_{(1,0),\vec{e}_{y}} \right.\right.\nonumber \\
&& + \left.\left. \hat{q}_{(1,1)} \right]^{2} + \left[ \hat{E}_{(0,0),\vec{e}_{x}} + \hat{E}_{(0,1),\vec{e}_{x}} - \hat{E}_{(1,1),\vec{e}_{x}} \right.\right.\nonumber \\
&& + \left.\left. \hat{q}_{(1,1)} + \hat{q}_{(1,0)} \right]^{2}  \right\rbrace .
\end{eqnarray}
Since Gauss' law affects the electric term only, the changes in the magnetic, mass, and kinetic contributions to the total Hamiltonians are trivial.
Note the natural appearance of the dynamical charges, which can be interpreted as sources of the electric field. 
Importantly, since $\hat{E}_{(1,0),\vec{e}_{x}}$, $\hat{E}_{(0,1),\vec{e}_{y}}$ and $\hat{E}_{(1,1),\vec{e}_{y}}$ are no longer dynamical variables, the corresponding raising and lowering operators become identities, leading to
\begin{eqnarray}\label{eq:single_plaq_link_ham}
\hat{H}_B =&& \frac{1}{2 g^2a^2} \left\lbrace \hat{U}_{(0,0),\vec{e}_{x}} \hat{U}_{(1,0),\vec{e}_{y}} \hat{U}_{(0,1),\vec{e}_{x}}^{\dag} \hat{U}_{(0,0),\vec{e}_{y}}^{\dag}  \right.\nonumber \\
&& + \left. \hat{U}_{(0,0),\vec{e}_{y}} \hat{U}_{(1,1),\vec{e}_{x}}^{\dag} \hat{U}_{(1,0),\vec{e}_{y}}^{\dag} + \hat{U}_{(1,1),\vec{e}_{x}} \right.\nonumber \\
&& + \left. \hat{U}_{(0,1),\vec{e}_{x}} \hat{U}_{(0,0),\vec{e}_{x}}^{\dag} + H.c. \right\rbrace , \nonumber \\
\hat{H}_M =&& m \left\lbrace \hat{\Psi}^\dag_{(0,0)} \hat{\Psi}_{(0,0)} - \hat{\Psi}^\dag_{(0,1)} \hat{\Psi}_{(0,1)} + \hat{\Psi}^\dag_{(1,1)} \hat{\Psi}_{(1,1)} \right. \nonumber \\
&& - \left. \hat{\Psi}^\dag_{(1,0)} \hat{\Psi}_{(1,0)} \right\rbrace, \nonumber \\
\hat{H}_K =&& \kappa \left\lbrace \hat{\Psi}^\dag_{(0,0)}\hat{U}_{(0,0),\vec{e}_{x}}\hat{\Psi}_{(1,0)} + \hat{\Psi}^\dag_{(0,0)}\hat{U}_{(0,0),\vec{e}_{y}}\hat{\Psi}_{(0,1)}  \right.\nonumber \\
&& + \left.  \hat{\Psi}^\dag_{(0,1)}\hat{U}_{(0,1),\vec{e}_{x}}\hat{\Psi}_{(1,1)} + \hat{\Psi}^\dag_{(1,0)}\hat{U}_{(1,0),\vec{e}_{y}}\hat{\Psi}_{(1,1)}   \right.\nonumber \\
&& + \left.  \hat{\Psi}^\dag_{(1,1)}\hat{U}_{(1,1),\vec{e}_{x}}\hat{\Psi}_{(0,1)} + \hat{\Psi}^\dag_{(0,1)}\hat{\Psi}_{(0,0)}   \right.\nonumber \\
&& + \left.  \hat{\Psi}^\dag_{(1,0)}\hat{\Psi}_{(0,0)} + \hat{\Psi}^\dag_{(1,1)}\hat{\Psi}_{(1,0)} + H.c.  \right\rbrace .
\end{eqnarray}
\subsection{An intuitive picture of the rotator and link operator relation}
\label{app:intuitive}
The representation of each rotator or string in terms of the link operators $\hat{E}_{\vec{n},\vec{e}_\mu}$ ($\mu =x,y$) can be interpreted as an analogue to Kirchhoff's loop law in electrical circuits. 
As shown in \figref{fig:theory}(a) the electric field $\hat{E}_{\vec{n},\vec{e}_x}$ ($\hat{E}_{\vec{n},\vec{e}_x}$) between vertices $\vec{n}$ and $\vec{n} + \vec{e}_x$ ($\vec{n} + \vec{e}_y$) is oriented along the positive $\vec{e}_x$ ($\vec{e}_y$) direction.  
Each rotator or string operator is then defined as the sum of the link contributions along it.
The sign of a contribution is positive, if the rotator or string is oriented with the positive field direction, or negative, if oriented opposingly.
For example, it holds that
\begin{align}
\hat{R}_1 &= E_{(0,0),\vec{e}_x} + (-1) (\hat{q}_{(1,0)} + \hat{q}_{(1,1)}) + \hat{E}_{(1,0),\vec{e}_y} \nonumber \\
&+ (- \hat{q}_{(1,1)}) + (-E_{(0,1),\vec{e}_x}) + (-E_{(0,0),\vec{e}_y}) \nonumber \\
& +  \hat{q}_{(1,0)},
\end{align}
where we explicitly marked opposing directions with a minus sign. 
Using the defining equations for the remaining operators and removing all redundant degrees of freedom yields an inverted form of \eqnref{eq:gaugefixing}, which reads
\begin{align}
  \hat{R}_1 & = -\hat{E}_{(0,1),\vec{e}_x}, \nonumber\\ 
    \hat{R}_2 & = -\hat{E}_{(0,1),\vec{e}_x} - \hat{E}_{(1,0),\vec{e}_y} - \hat{q}_{(1,1)}, \nonumber\\
    \hat{R}_3 & = -\hat{E}_{(1,1),\vec{e}_y} \nonumber\\
    \hat{R}_x & = -\hat{E}_{(0,0),\vec{e}_x} + \hat{E}_{(0,1),\vec{e}_x} + \hat{q}_{(1,0)} + \hat{q}_{(1,1)}, \nonumber\\
    \hat{R}_y & = \hat{E}_{(0,1),\vec{e}_y} + \hat{E}_{(1,1),\vec{e}_y}.
\end{align}

\subsection{Gauge field creation in the rotator and string picture}
\label{sec:detailsKinetic}
\begin{figure*}[ht!]
\includegraphics[width=2\columnwidth]{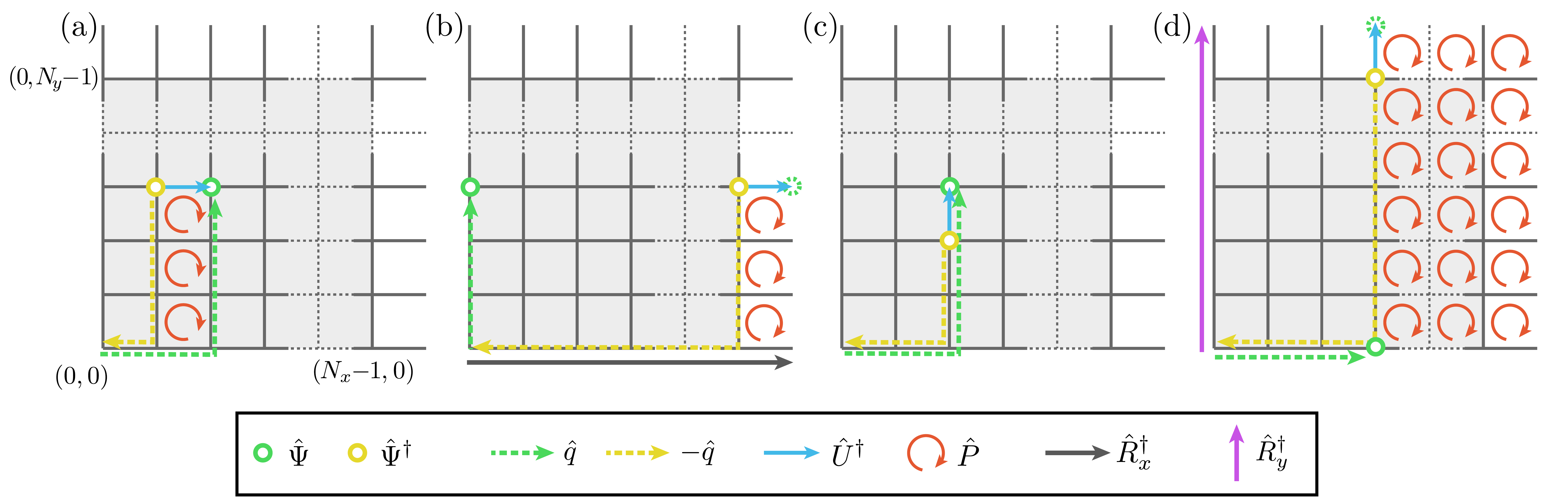}
\caption{\textbf{Pair creation on a periodic two-dimensional lattice.} The four panels describe the creation of a gauge field (blue arrow) in $x$ direction [(a) and (b)], and in $y$ direction [(c) and (d)]. The particles (green and yellow circles) imply the creation of electric fields on the links marked with the corresponding dashed arrows. Only in (c) these contributions are enough to create the gauge field while maintaining gauge-invariance in all other links. Otherwise, the gauge field has to be created by annihilating the plaquette operators marked with red circular arrows, which also counteracts the action of the particles' charges. Note, that the field on a link is effectively unchanged if it is modified by an equal number of arrows in both directions, which in (b) and (d) requires the introduction of the strings (grey and pink) when the particles are created on the boundary condition of the lattice.}
\label{fig:kinetic}
\end{figure*}

In the following, we illustrate the formulation of the kinetic Hamiltonian \eqnref{eq:KineticContrGen} for a general two-dimensional plane in the rotator and string formulation. 
In particular, the mapping involves transformations of the type
\begin{equation}\label{eq:general_rule_kin}
\hat{\psi}^{\dagger}_{\vec{n}}\hat{U}^{\dagger}_{\vec{n},\vec{e}_\mu}\hat{\psi}_{\vec{n}+\vec{e}_{\mu}} \mapsto \hat{\psi}^{\dagger}_{\vec{n}}f_{\vec{n},\vec{e}_{\mu}}\left(\left\lbrace \hat{P}, \hat{P}^{\dagger}\right\rbrace \right)\hat{\psi}_{\vec{n}+\vec{e}_{\mu}},
\end{equation}
where $\mu = x,y$ and $f_{\vec{n},\vec{e}_{\mu}}$ is a function of both rotator and string operators. Depending on the particle pair to be created or annihilated, there are four distinct rules for building up the functions $f_{\vec{n},\vec{e}_{\mu}}$. These are shown in the four corresponding panels displayed in \figref{fig:kinetic}, where we use a yellow (green) circle to indicate an (anti)fermion, and light blue arrows for the corresponding link in which the electric field is required to raise [$\hat{U}^{\dagger}_{\vec{n},\vec{e}_{\mu}}$ in \eqnref{eq:general_rule_kin}]. Employing the notation presented in the legend of \figref{fig:kinetic}, we can directly build the functions $f_{\vec{n},\vec{e}_{\mu}}$ in \eqnref{eq:general_rule_kin} and recover the rules presented in \eqnref{eq:kin}. 

As an example, consider \figref{fig:kinetic}(a), which describes all cases in which a pair is created at locations $\vec{n}$ and $\vec{n}+\vec{e}_x$, with $n_x \neq N_x-1$. If $n_y=0$, the electric field is automatically corrected by the chosen charge strings that ensure Gauss' law (dotted green and yellow lines in \figref{fig:kinetic}; see \secref{app:Reparametrisation} and \secref{sec:generalTorus}), implying $f_{(n_x,0),\vec{e}_{x}} = \eins$. Otherwise ($n_y \neq 0$), to increase the electric field $\hat{E}_{\vec{n},\vec{e}_x}$ on the link between the two charges, we lower the rotator $\hat{R}_{(n_x,n_y-1)}$ below by applying $\hat{P}_{(n_x,n_y-1)}$. This, however, affects the electric fields on all other links forming $\hat{R}_{(n_x,n_y-1)}$. While the vertical ones are taken care of by the charge strings, the bottom link is not (unless $n_y = 1$). As graphically explained in \figref{fig:kinetic}(a), to increase only the desired link, we can lower all rotators $\hat{R}_{(n_x,r_y)}$ with $r_y = 0,...,n_y-2$ below, yielding $f_{(n_x,n_y),\vec{e}_{x}} = \prod_{r_y=0}^{n_y-1} \hat{P}_{(n_x,r_y)}$. 

The remaining panels in \figref{fig:kinetic} further illustrate the cases of pairs that are connected by vertical links and pairs that are created on links closing the periodic boundary conditions, where we require the strings $\hat{R}_{x,y}$. By following the same procedure above, it is then possible to determine the functions $f_{\vec{n},\vec{e}_{\mu}}$ for all allowed choices of $\vec{n}$ and $\mu = x,y$, yielding \eqnref{eq:kin}.

\section{Diagonalisation of the magnetic gauge field Hamiltonian}
\label{app:fourier}

The magnetic Hamiltonian $\hat{H}_B$ is composed of the lowering and raising operators $\hat{P}_j$, $\hat{P}_j^\dagger$, $j=1,2,\dots,N-1$, where $N$ is the total number of plaquettes. 
In the $\mathbb{Z}_{2L+1}$ group, these operators are the so-called cyclant matrices, which can be diagonalised exactly. 
Before truncation, the lowering operators are defined according to \eqnref{eq:def_raising_ZN},
\begin{eqnarray}
\hat{P}_j = \ketbra{L_j}{-L_j} + \sum_{r_j=-L_j+1}^{L_j} \ketbra{r_j-1}{r_j}.
\end{eqnarray}
For the sake of simplicity, we drop the index $j$ and note that the procedure is equivalent for all subsystems.
The spectrum of the lowering operators is
\begin{eqnarray}
\omega_k = e^{-i\frac{2\pi}{2L+1}k},
\end{eqnarray}
while the corresponding eigenvectors are given by
\begin{eqnarray}
v_k = \frac{1}{\sqrt{2L+1}}(\omega_k^{L}, \omega_k^{L-1}, \dots , \omega_k^{-L})^{\t{T}},
\end{eqnarray}
with $k=-L,\dots,L$. Hence, $\hat{U}$ is diagonalised by the matrix
\begin{eqnarray}
V^\dagger &=& (v_{-L}, v_{-L+1}, \dots , v_{L}) \nonumber \\
&=& \frac{1}{\sqrt{2L+1}} \sum_{\mu,\nu=-L}^{L} \t{e}^{-i\frac{2\pi}{2L+1}\mu\nu}\ketbra{\mu}{\nu} \nonumber \\
&\equiv & \hat{\mathcal{F}}_{2L+1}^\dagger,
\end{eqnarray}
which is the discrete Fourier transform.
Hence, it is straightforward to show that
\begin{eqnarray}
\hat{\mathcal{F}}_{2L+1}\hat{P}^{\gamma}\hat{\mathcal{F}}_{2L+1}^{\dag} 
= \sum_{r=-L}^{L} \e^{-i\frac{2\pi}{2L+1}\gamma r}\ketbra{r}{r},
\end{eqnarray}
where $\gamma \in \mathbbm{Z}$. 
Moreover, for any $N-1 \geq J \in \mathbb{N}$, we have that
\begin{eqnarray}
\hat{\mathcal{F}}_{2L+1}\Bigg[\bigotimes_{j=1}^J&\hat{P}_j^{\gamma}&\Bigg]\hat{\mathcal{F}}_{2L+1}^{\dag} \nonumber \\
&=&\sum_{\vec{r}=-\vec{L}}^{\vec{L}} \e^{-i\frac{2\pi}{2L+1}\vec{\gamma}\vec{r}}\ketbra{\vec{r}}{\vec{r}}.
\end{eqnarray}
Here, we use $\vec{r}=(r_1, r_2,\dots,r_J)^\t{T}$ and $\vec{\gamma}=(\gamma_1,\gamma_2,\dots,\gamma_J)^\t{T}$, while we waived to denote that the Fourier transform is now understood as the product of the Fourier transforms in the separate $N-1$ spaces.
Note that, in particular $(\hat{P}^\gamma)^\dagger = \hat{P}^{-\gamma}$ and therefore:
\begin{eqnarray}
&\hat{\mathcal{F}}_{2L+1}&\bigg[\bigotimes_{j=1}^J \hat{P}_j^{\gamma}\pm\bigotimes_{j=1}^J \hat{P}_j^{-\gamma}\bigg]\hat{\mathcal{F}}_{2L+1}^{\dag} \nonumber \\ 
&=& 2\sum_{\vec{r}=-\vec{L}}^{\vec{L}} \begin{cases} \cos(\frac{2\pi \vec{r} \vec{\gamma}}{2L+1})\ketbra{\vec{r}}{\vec{r}} &\; \t{for}\,+\\
-i\sin(\frac{2\pi \vec{r} \vec{\gamma}}{2L+1})\ketbra{\vec{r}}{\vec{r}}&\; \t{for}\,-
\end{cases}.
\end{eqnarray}
Let us finally remark that these relations hold interchangeably for $\hat{P}$ in the rotator formalism and $\hat{U}$ for the electric fields as introduced in \secref{sec:QEDin2D}.

\section{Asymptotic behaviour of the ground state expectation value of $\Box$}
\label{app:asymptotic}
In this appendix, we describe the behaviour of the ground state expectation value $\langle \Box \rangle$ both in the electric and magnetic representation. We look at the extremal regimes and study truncation effects.

We indicate the average value of the observable $\Box$ in the electric (magnetic) representation with $\langle\Box\E\rangle$ ($\langle\Box\B\rangle$).
Recall that $\Box = -g^2 \hat{H}_B/4$, and consider the electric representation first. 
Here, $\hat{H}_B\E$ is composed of the Hermitian operators $\hat{P}_i + \hat{P}_i^\dagger$ ($i = 1,2,3$), and of $\hat{P}_{1}\hat{P}_{2}\hat{P}_{3} + (\hat{P}_{1}\hat{P}_{2}\hat{P}_{3})^\dagger$, where the latter is the replacement for $\hat{P}_{4}+ \hat{P}_{4}^\dagger$ (see \secref{sec:single_plaq_ham}).
Hence, we can bound the spectrum of $-\hat{H}_B$ as $-\bra{\psi}\hat{H}_B\ket{\psi}\leq 4 \bra{\psi}(\hat{P}_i + \hat{P}_i^\dagger)\ket{\psi}/(2g^2) = 2\lambda_{\t{max}}/g^2$.
Importantly, this holds for all $i$ since the operators describe the rotators which represent identical systems.
In the regime where $g \ll 1$, $-2\lambda_{\t{max}}/g^2$ corresponds to the ground state energy and consequently $\lambda_\t{max}/2$ to $\langle\Box\E\rangle$.
Now, as long as the Hamiltonian is not truncated, i.e. $l = L$, we have that $\hat{P}_i + \hat{P}_i^\dagger$ is a circulant matrix \cite{Gray2006toeplitz} and thus its eigenvalues are given as
\begin{equation}
\xi_j = 2 \cos\left(\frac{2\pi j}{2L+1}\right), \quad j = 0,1,\dots,2L.
\end{equation}
Consequently, $\lambda_\t{max}=2$ and thus $\langle\Box\E (g=0)\rangle = 1$.
However, as soon as the Hamiltonian is truncated, $\hat{P}_i + \hat{P}_i^\dagger$ is a tridiagonal Toeplitz matrix \cite{Gray2006toeplitz} with eigenvalues
\begin{equation}
\xi_j = 2 \cos \left(\frac{\pi (j+1)}{2l+2}\right), \quad j = 0,1,\dots,2l.
\end{equation}
This yields $-\bra{\psi}\hat{H}_B\ket{\psi} \leq 4\cos\left(\frac{\pi}{2l+2}\right)$, which results in a monotonic increase of $\langle\Box\E(g=0)\rangle$ with respect to an increase of $l$.
The continuous limit is found for $l\rightarrow\infty$. 
Since the electric representations is based on the truncated $U(1)$ group, we conclude that for $g \ll 1$ it approximates the true $U(1)$ value from below. 

Consider now the magnetic representation. The curves in \figref{fig:plaquette_expectation} suggest that $\langle\Box\B\rangle$ is monotonically decreasing over the whole range of $g$ when $l$ is increased. In the following, we qualitatively motivate this behaviour in the strong coupling regime. When $g \gg 1$, $\langle\Box\B\rangle$ can be understood as a Riemann sum of the discrete eigenvalues of $\hat{H}_B\B$, weighted with the probabilities corresponding to the different basis states $\ket{\vec{r}}$. 
Assume for now $l=L$. Then, the ground state emerges as the equal superposition of all states (as in \eqnref{eq:GS_magnetic_on_electric_side}, where we considered the electric representation for $g \ll 1$). 
The weights of the Riemann sum are thus all equal to $(2L+1)^{-3}$, meaning that
\begin{eqnarray}
\langle\Box\B\rangle &=& \sum_{\vec{r}=-\vec{L}}^{\vec{L}} \frac{1}{(2L+1)^3} \left[\sum_{i}\cos\left(\frac{2\pi}{2L+1}r_i\right) \right. \nonumber \\
&+& \left.\cos\left(\frac{2\pi}{2L+1}(r_1+r_{2}+r_{3})\right) \right] = 0.
\label{eq:PlaquetteAverageMagnetic}
\end{eqnarray}
Under the assumption that the distribution of the ground state's coefficients $p\B(g\gg 1,\vec{r}, L)$ (see \secref{sec:phenomenology}) remains uniform in $\vec{r}$ for $l<L$, one can show that this value is larger than zero for any truncation. In particular,
\begin{eqnarray}
\langle\Box\B\rangle \sim  3(2l+1)^3 \sin \left(\frac{2 \pi  (L-l)}{2 L+1}\right) \underset{l\rightarrow L}{\longrightarrow} 0^{+}.
\end{eqnarray}
However, the distribution $p\B(g \gg 1,\vec{r}, L)$ is not uniform when $l<L$, but remains center-symmetric.
We numerically show in \appref{app:truncation} that removing higher levels increases the amplitudes of states with low values of $|\vec{r}|$, and hence enhances positive contributions of the Riemann sum.
In other words, truncation not only removes the negative contributions stemming from large values of $|\vec{r}|$ [they appear for $r_i > (2L+1)/4$, see \eqnref{eq:PlaquetteAverageMagnetic}], but also emphasises positive contributions. While the first process is analogous to the decrease of the spread when decreasing $g$, the second is solely originating from the truncation and we therefore conclude that for any $L$ the approximated value of $\langle\Box\B\rangle$ is always larger than the one obtained from a hypothetical exact diagonalisation employing $U(1)$.
Qualitatively, this effect emerges from the removal of the cyclic property present in the lowering operator $\hat{P}$ (see \eqnref{eq:def_raising_ZN} and \appref{app:truncation}).

\section{Truncation effects in the strong coupling regime}
\label{app:truncation}
\begin{figure}[t]
\centering 
\includegraphics[width=.9\columnwidth]{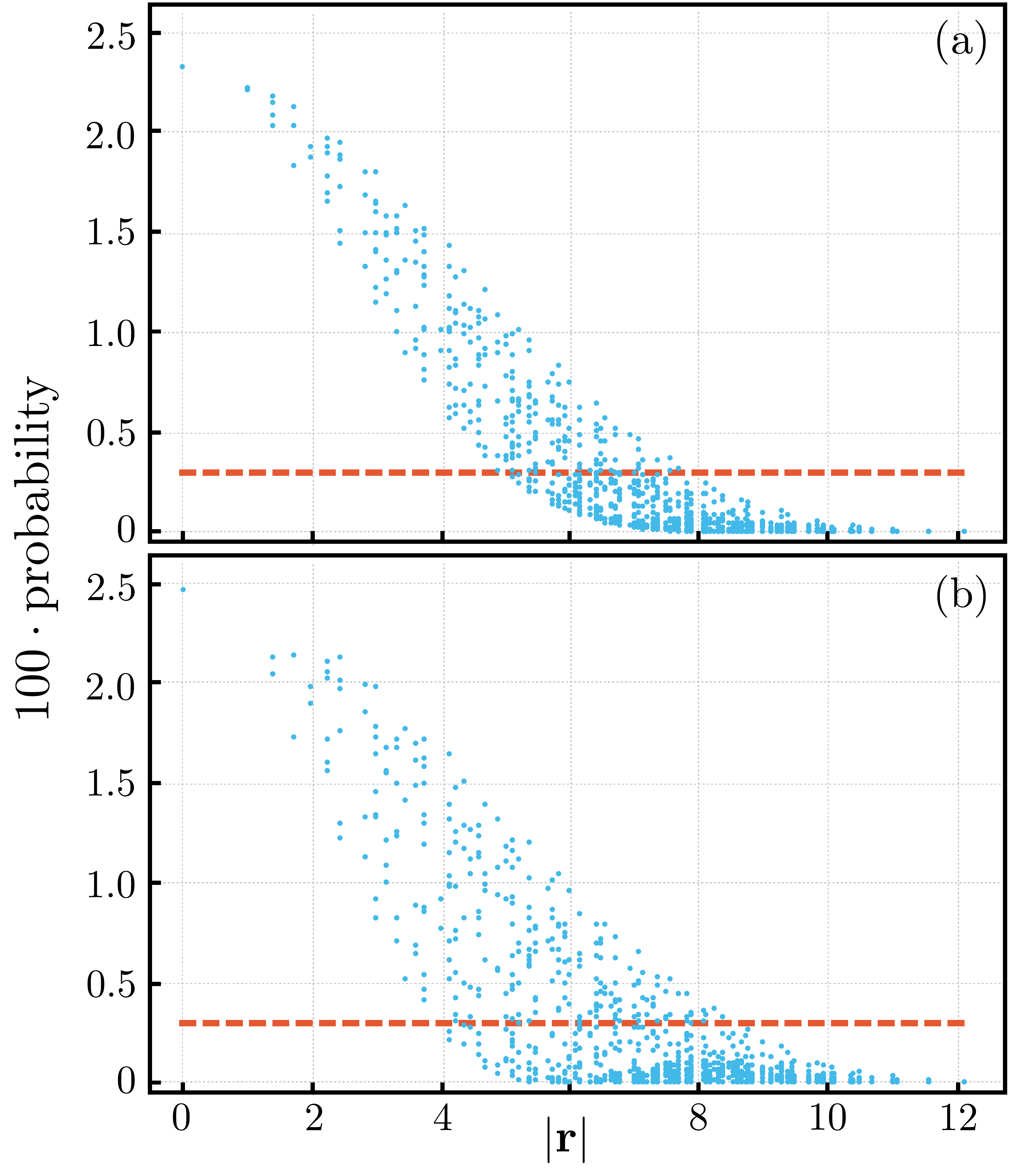}
\caption{\textbf{Transformation of the ground state distribution after truncation.} Both panels correspond to $l=7$ and $L=8$, the x-axis labels the amplitudes belonging to $\ket{\vec{r}}$. Compared to (a), (b) shows the effect when only the elements corresponding to the cyclic property of $\mathbb{Z}_{2L+1}$ are removed.}
\label{fig:truncationPerturbation}
\end{figure}
In this appendix, we sketch the treatment of the truncation of the cyclic $\mathbb{Z}_{2L+1}$ group.
We consider the strong coupling regime, but employ the magnetic representation of the Hamiltonian.
In the limit $g\rightarrow\infty$, the electric term, which is composed of operators $\hat{P}_j$, $j = 1,2,3$ [see \eqnref{eq:def_raising_ZN}], is dominant. 
As such, we ignore the magnetic Hamiltonian in the following and employ the following ansatz to obtain the truncated electric field Hamiltonian
\begin{eqnarray}
H_{E,\t{truncated}}\B = \hat{H}_{E,\t{untruncated}}\B - \hat{V}.
\label{eq:perturbation_ansatz}
\end{eqnarray}
We hence define the operator $\hat{V}$ to study the effects of the truncation.

In the untruncated $\mathbbm{Z}_{2L+1}$ formulation we can decompose any $\hat{P}_j$ into four terms,
\begin{eqnarray}
\hat{P}_j &=& \ketbra{L_j}{-L_j}  + \sum_{r_j=l+1}^L \ketbra{r_j-1}{r_j} \nonumber \\ 
&&+ \sum_{r_j=1-L}^{-l} \ketbra{r_j-1}{r_j} \nonumber \\
&&+ \sum_{r_j=1-l}^{l} \ketbra{r_j-1}{r_j} \nonumber \\
&\equiv & \hat{V}^\prime_j + \hat{P}^\prime_j,
\label{eq:U_decomp}
\end{eqnarray}
where $\hat{P}^\prime_j = \sum_{r_j=1-l}^{l} \ketbra{r_j-1}{r_j}$ and $\hat{P}^\prime_j$ is the rest. Notice that $\hat{P}^\prime_j$ is the truncated operator as defined in \eqnref{eq:def_raising_truncation}. 
We now have to collect all contributions from $\hat{V}^\prime_j$ in the electric Hamiltonian $\hat{H}_E\B$ of \eqnref{eq:HE_in_B}. In particular, $\hat{V}$ can be found from $\hat{H}_E\B$ by applying the rules
\begin{eqnarray}
P_j + P_k^\dagger &\mapsto& \hat{V}^\prime_j + (\hat{V}_k^\prime)^\dagger, \nonumber\\
P_j P_k &\mapsto& \hat{V}^\prime_j (\hat{V}_k^\prime)^\dagger + \hat{V}^\prime_j (\hat{P}_k^\prime)^\dagger + \hat{P}^\prime_j (\hat{V}_k^\prime)^\dagger.
\end{eqnarray}
Note that $\hat{V}$ has as well a diagonal contribution given by
\begin{eqnarray}
\sum_{j=1}^3 \sum_{r_j=l+1}^{L} \ketbra{r_j}{r_j} + \ketbra{-r_j}{-r_j}, 
\end{eqnarray}
stemming from the $\nu=0$ terms in \eqnref{eq:HE_in_B}.

The result of a numerical procedure to find the elements of ground state is shown in \figref{fig:truncationPerturbation}(a).
We denote the amplitudes of a state obtained with the untruncated Hamiltonian with the red dashed line. 
Clearly, the truncation shifts population from the high $|\vec{r}|$ states to lower ones. 
Note that each $|\vec{r}|$ can appear multiple times.
A crucial fact is shown in \figref{fig:truncationPerturbation}(b), where we removed only the cyclic elements from the operators $P_j$, i.e., the first terms in \eqnref{eq:U_decomp}.
This cyclic property is the reason why the distributions $p\B(\vec{r})$ of the ground state's coefficients (see \secref{sec:phenomenology}) are uniform in the untruncated case. Hence, their removal has a large impact on the derived results.

\section{Numerical determination of $L_{\rm opt}$}
\label{app:optimalL}

\begin{figure}[t]
	\centering 
	\includegraphics[width=.9\columnwidth]{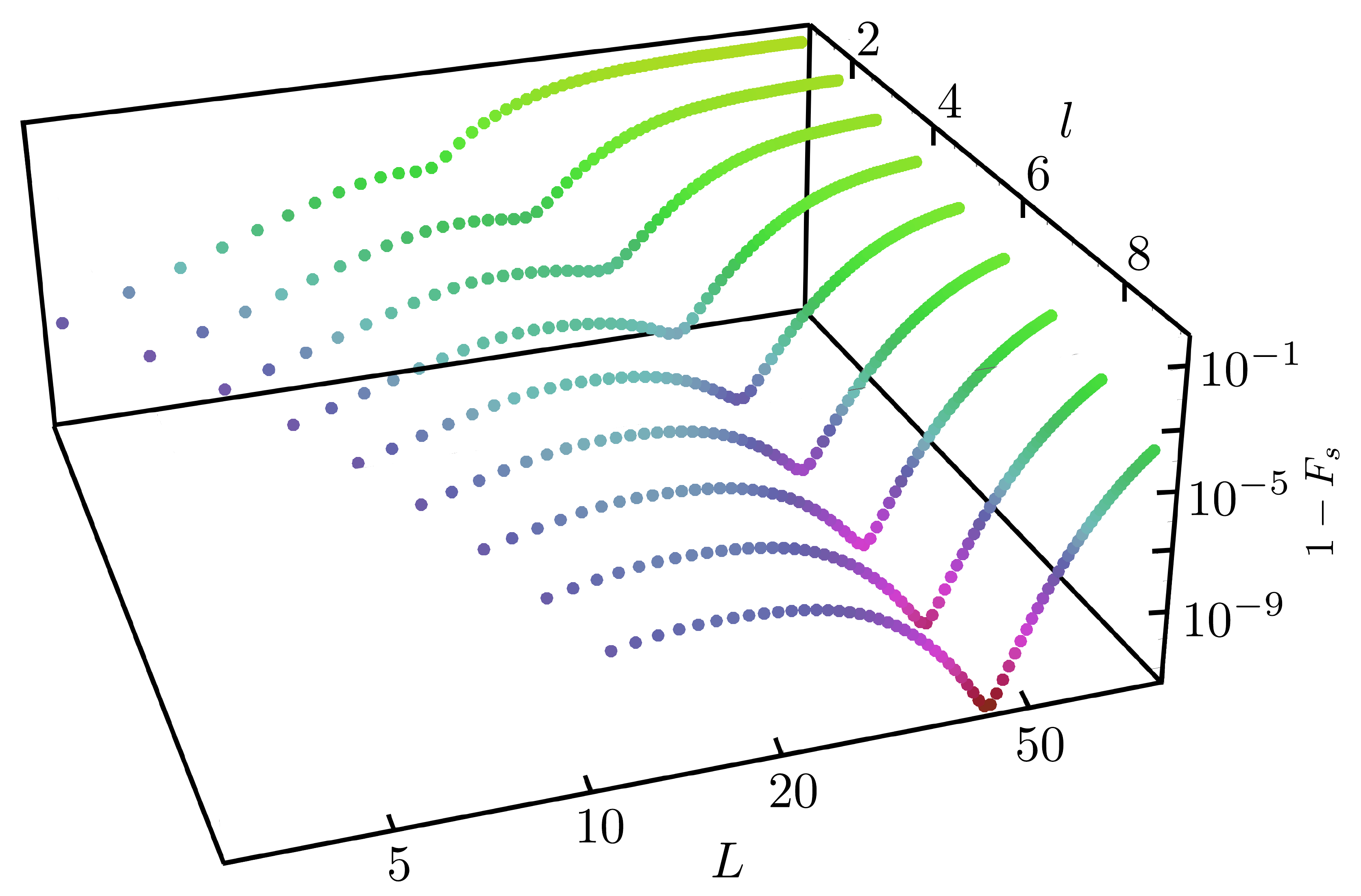}
	\caption{\textbf{Sequence infidelity} for $g^{-2} = 100$ and $l=2, 3, \dots, 10$. The minimum corresponds to the value of $L$ with the best compromise between available domain and resolution. Note that this minimum is not always the global one.}
	\label{fig:Lopt}
\end{figure}
The sequence fidelity calculated in \secref{sec:fidelity} for the ground state of the pure gauge QED Hamiltonian involves and optimization over $L$.
We plot the sequence infidelity varying the parameters $l$ and $L$ for $g^{-2} = 100$ in \figref{fig:Lopt}. 
For any value of $l$, a kink is clearly visible, corresponding to $L=L_{\rm opt}$. 
Notice that this kink does not always correspond to a global minimum, as can be seen from the points characterised by $l=2$. 
As discussed in the main text, this is the signature of the freezing effect. 
In fact, the global minimum is found for $L=l+1=3$ (i.e. its minimal value), where the resolution is insufficient for both $l=1$ and $l=2$ to capture the distribution of the untruncated ground state. 
By increasing $g^{-2}$, the position of the kink is shifted to higher values of $L$. In particular, $L_{\rm opt}$ takes its minimal allowed value in the strong coupling regime, and starts to increase at $g^{-2} \approx 5$ [see \figref{fig:state_convergence}(d)]. This follows from the fact that, approaching the weak coupling regime, the distribution of the untruncated $U(1)$ ground state becomes more and more localised, and the tails less important.
Note that the value $L_{\rm opt}$ can be determined by a greed search, starting at the lowest allowed value of $L$.

\end{document}